\documentclass[aps,pre,twocolumn,superscriptaddress,10pt]{revtex4}
\usepackage[dvips]{graphics}
\usepackage{graphicx}
\usepackage{amsfonts}
\usepackage{amssymb}
\usepackage{amsmath}

\begin{document}

\title{Spinor Bose-Einstein condensate flow past an obstacle}

\author{A.\ S.\ Rodrigues}
\affiliation{Departamento de F\'{\i}sica/CFP, Faculdade de Ci\^{e}ncias, Universidade do Porto, R. Campo Alegre,
687 - 4169-007 Porto, Portugal}
%\email{asrodrig@fc.up.pt}

\author{P.\ G.\ Kevrekidis }
\affiliation{Department of Mathematics and Statistics, University of Massachusetts,
Amherst MA 01003-4515, USA}
%\email{kevrekid@math.umass.edu}

\author{R.\ Carretero-Gonz\'alez}
\affiliation{Nonlinear Dynamical Systems Group\footnote{\texttt{URL}: http://nlds.sdsu.edu}%
, Department of Mathematics and Statistics, and Computational Science
Research Center, San Diego State University, San Diego CA, 92182-7720, USA}

\author{D.\ J.\ Frantzeskakis}
\affiliation{Department of Physics, University of Athens, Panepistimiopolis, Zografos, Athens 157 84, Greece}

\author{P.\ Schmelcher}
\affiliation{Theoretische Chemie, Physikalisch-Chemisches Institut, Im Neuenheimer Feld 229,
Universit\"at Heidelberg, 69120 Heidelberg, Germany}
\affiliation{Physikalisches Institut, Universit\"at Heidelberg, Philosophenweg 12, 69120 Heidelberg, Germany}

\author{T.\ J.\ Alexander}
\affiliation{Nonlinear Physics Center, Research School of Physical Sciences and
Engineering, Australian National University, Canberra ACT 0200,
Australia}

\author{Yu.\ S.\ Kivshar}
\affiliation{Nonlinear Physics Center, Research School of Physical Sciences and
Engineering, Australian National University, Canberra ACT 0200,
Australia}

\begin{abstract}
We study the flow of a spinor ($F=1$) Bose-Einstein condensate
in the presence of an obstacle. We consider the cases of ferromagnetic and polar
spin-dependent interactions and find that the system demonstrates two speeds of sound that are
identified analytically. Numerical simulations reveal the nucleation of macroscopic nonlinear 
structures, such as dark solitons and vortex-antivortex pairs, as well as vortex rings in one- 
and higher-dimensional settings respectively, when a localized defect (e.g., a blue-detuned 
laser beam) is dragged through the spinor condensate at a speed larger than the second critical speed.
\end{abstract}

\maketitle

\section{Introduction}

Over the last decade, we have seen an enormous growth of interest and a related diversification of the physics of atomic Bose-Einstein condensates (BECs) \cite{book1,book2}. A significant aspect of this ever-expanding interest is the intense study of macroscopic nonlinear excitations, such as solitons and vortices, which can arise in BECs~\cite{ourbook}. In fact, the emergence of such macroscopic coherent
structures in the many-body state of the system establishes a close connection between
BECs and other branches of physics, such as e.g., optics and the physics of nonlinear waves.
Within this interface of atomic and nonlinear wave physics, in recent years there has been
an increasing focus on the study of multi-component BECs~\cite{ourbook},
and particularly {\it spinor condensates}~\cite{ket1,cahn}. The latter have been realized with the help of far-off-resonant optical techniques for trapping ultracold atomic gases~\cite{ket0}
which, in turn, allowed the spin degree of freedom to be explored (previously frozen in magnetic traps).
This relatively recent development has given rise to a wealth of multi-component phenomena,
including the formation of spin domains~\cite{spindomain} and spin textures~\cite{spintext},
spin-mixing dynamics~\cite{pulaw}, dynamic fragmentation~\cite{yousun}, and the dynamics of quantum phases~\cite{yiyou}.
At the same time, macroscopic nonlinear structures that may arise in
spinor BECs have also been investigated. Such structures include bright~\cite{wad1,boris,zh}, dark~\cite{wad2}, and
gap solitons~\cite{dabr}, as well as more elaborate complexes, such as bright-dark solitons~\cite{ourbd}
and domain-walls~\cite{ourdw}.

A relevant direction that has been of particular interest concerns the study
of the breakdown of superfluidity and the concomitant generation of excitations in BECs.
In particular, much experimental and theoretical effort has been devoted to the
understanding of important relevant concepts, such as the critical velocity
introduced by Landau, sound waves and the speed of sound, and the emergence of vortices
and solitons~\cite{book1,book2}. From the theoretical point of view, the Gross-Pitaevskii (GP) equation has been used to study the flow of a BEC around an obstacle or, equivalently, the effect of dragging a localized potential
(such as a blue-detuned laser beam) through a BEC. In this context, it has
been predicted theoretically~\cite{hakim,radouani,george,our1d} and recently observed
experimentally~\cite{engels} that when the speed of the ``localized defect'' exceeds
a critical speed, then dark solitons are formed in quasi one-dimensional (1D) condensates.
On the other hand, in higher-dimensional [e.g., quasi two-dimensional (2D)] settings,
theoretical studies~\cite{adams} have shown that a similar procedure leads to the formation of
vortices (or more precisely to vortex-antivortex pairs, due to the conservation of total topological charge).
Importantly, experimental consequences of this procedure, such as an onset of heating and dissipation were
monitored experimentally~\cite{kettexp}. Other relevant theoretical works include studies of
the breakdown of superfluidity, the onset of dissipation, and the associated Landau criterion~\cite{pavloff}. More recently, dragging of an obstacle in a two-component BEC was studied in Ref.~\cite{susanto}. In this latter study
it was established that two distinct ``speeds of sound'' arise and the form of the ensuing nonlinear
structures (e.g., dark-dark or dark-anti-dark soliton pairs in 1D, and vortex-vortex or vortex-lump pairs in 2D) depend on how the value of the obstacle speed compares to the values of the critical speeds.

In this paper, we consider the dragging of a localized defect through an $F=1$ spinor condensate
with repulsive spin-independent interactions and either ferromagnetic or
anti-ferromagnetic (polar) spin-dependent interactions. In the framework of
mean-field theory, this system is described by a set of coupled
GP equations for the wavefunctions of the three hyperfine components.
A key question that emerges in this $F=1$ spinor BEC setting
is how many critical speeds may be available. A naive
count based on the three-component nature of the system
(and by analogy to the two-component setting bearing two such
critical speeds) would suggest the possibility for three distinct
critical speeds. However, as we illustrate below, an explicit calculation
reveals that there exist only two such critical speeds in the system,
due to the particular nature of the nonlinearity. Moreover, our numerical simulations illustrate that the crossing of the lower of the two critical speeds does not appear to lead to the formation
of nonlinear excitations. On the other hand, for defect speeds larger
than the second critical speed, our simulations illustrate that
dark solitons emerge in the 1D setting, vortex-antivortex pairs in the 2D setting, and
vortex rings are shown to arise in the fully three-dimensional (3D) setting.

The presentation of our results is structured as follows.
In Sec.~II, we develop an analytical approach for computing the
relevant critical speeds, by generalizing to the spinor setting the
arguments of Ref.~\cite{hakim}. Then, in Sec.~III, we numerically test
the relevant predictions in 1D, 2D and 3D settings. Finally, in Sec.~IV, we summarize our findings and
point to some important remaining questions along this vein of research.

%%%%%%%%%%%%%%%%%%%%%%%%%%%%%%%%%%%%%%%%%%%%%%%%%%%%%%%

\section{Model and its analysis}

In our analytical approach, we will consider a quasi-1D spinor $F=1$ BEC with
repulsive spin-independent interactions. In the framework of mean-field theory, this
system can be described by the following normalized GP equations \cite{ourbd,ourdw}:
\begin{eqnarray}
 i\partial _{t}\psi _{\pm 1} &=& H_{\mathrm{0}} \psi_{\pm 1}
+ r \left[(|\psi _{\pm 1}|^{2}+|\psi _{0}|^{2}-|\psi_{\mp 1}|^{2})\psi_{\pm 1} \right]
\notag
\\&&+ r \psi_{0}^{2}\psi_{\mp 1}^{\ast},
\label{eq012}
\\[1.0ex]
i\partial_{t}\psi _{0} &=&H_{\mathrm{0}}\psi _{0}+
r \left[ (|\psi_{-1}|^{2}+|\psi _{+1}|^{2})\psi _{0}\right]
\notag
\\&&+ 2r \psi _{-1}\psi _{0}^{\ast }\psi_{+1},
\label{eq0}
\end{eqnarray}
where $H_{\mathrm{0}}\equiv -(1/2)\partial_{x}^{2}+ V(x;t)+n_{\mathrm{tot}}$, while
$n_{\mathrm{tot}}=|\psi_{-1}|^{2}+|\psi _{0}|^{2}+|\psi _{+1}|^{2}$ is the total density and
$V(x;t)$ is the external potential. The latter is assumed to take the following form:
\begin{eqnarray}
V(x;t)=\frac{1}{2} \Omega^2 x^2 + V_0\, \exp (-a (x-s t)^2).
\label{eq4}
\end{eqnarray}
The first term in the right-hand side of Eq.~(\ref{eq4}) represents a typical harmonic trapping
potential of normalized strength $\Omega$, while the second term accounts for a localized
repulsive potential (e.g., a blue-detuned laser beam), of strength $V_0$ and width $a^{-1}$,
that is dragged through the condensate at speed $s$. Note that our analytical results
will be obtained below for the case of $\Omega=0$ (which still contains the fundamental
phenomenology), but
%will also be
were also tested in the numerical simulations for $\Omega \neq 0$
(and
%will be
were found to persist in the latter case). Finally, the parameter $r$ in
Eqs.~(\ref{eq012})-(\ref{eq0})
expresses the normalized spin-dependent interaction strength
defined as $r =(a_{2}-a_{0})/(a_{0}+2a_{2})$, where $a_{0}$ and $a_{2}$ are the
$s$-wave scattering lengths in the symmetric channels with total spin of the
colliding atoms $F=0$ and $F=2$, respectively. Note that $r<0$ and $r>0$ correspond,
respectively, to ferromagnetic and polar spinor BECs. In the relevant cases
of $^{87}$Rb and $^{23}$Na atoms with $F=1$, this parameter takes values $%
r =-4.66\times 10^{-3}$ \cite{kemp} and $r =+3.14\times 10^{-2}$
\cite{greene}, respectively, i.e., in either case, it is a small parameter
in Eqs.~(\ref{eq012})-(\ref{eq0}).

We now seek uniform stationary solutions of the GP Eqs.~(\ref{eq012})-(\ref{eq0}) (with $V = 0$)
in the form
\begin{eqnarray}
\notag
\psi_1&=&A \exp(-i \mu_{+1} t)\exp(i\theta_{+1}), \\[1.0ex]
\notag
\psi_0&=&B \exp(-i \mu_0 t)\exp(i\theta_0), \\[1.0ex]
\notag
\psi_{-1}&=&C \exp(-i \mu_{-1} t)\exp(i\theta_{-1}),
\end{eqnarray}
where $A$, $B$, $C$
and $\theta_j$ (with $j\in\{-1,0,+1\}$) represent, respectively, the amplitudes and phases
of the hyperfine components, and $\mu_j$ are their chemical potentials.
In our analysis below we will assume that $A \neq 0$, $B \neq 0$, $C \neq 0$,
as that will provide us with genuinely spinor (i.e., three-component) states;
otherwise the system is reduced to a lower number of components.  In fact,
the analysis for the one-component case has been carried out
in Ref. \cite{hakim}, while
in the two-component case considerations analogous to the ones that we will present below
have been put forth in Ref. \cite{susanto}. Under the above genuinely three-component assumption,
we substitute the stationary solutions into the GP Eqs.~(\ref{eq012})-(\ref{eq0}) and
obtain the following set of equations:
\begin{eqnarray}
\mu_{+1} &=& n_{\mathrm{tot}} + r (A^2+B^2-C^2) + p r \frac{B^2 C}{A},
\notag
\\
\mu_0 &=&n_{\mathrm{tot}} + r (A^2 + C^2) + 2 p r A C,
\notag
\\
\mu_{-1} &=&n_{\mathrm{tot}} + r (C^2+B^2-A^2) + p r \frac{B^2 A}{C},
\notag
\end{eqnarray}
where $n_{\mathrm{tot}}=A^2+B^2+C^2$.

In the above expressions phase matching conditions were used, 
as is usual when one has parametric interactions:
these read $2\mu_0=\mu_{+1}+\mu_{-1}$ for the chemical potentials, and
$\Delta \theta=2\theta_0 -(\theta_{+1}+\theta_{-1}) =0$ or $\pi$
for the relative phase between the hyperfine components \cite{ostro,ourdw}.
The factor $p \equiv \pm 1$  on the last term of each of the above
equations results from considering
$\Delta \theta=0$
or $\pi$, respectively. In the case where the three chemical potentials $\mu_j$
are different, it can be found
that it is not possible to satisfy the above assumption
that each of the amplitudes $A$, $B$, $C$ should be nonzero.
Hence, we will hereafter
focus on the case of $\mu_{+1}=\mu_0=\mu_{-1} \equiv \mu$.
In the latter case, it is straightforward to algebraically manipulate the equations
and find that
there exist only two classes of possible stationary solutions with
a free parameter (for a given $\mu$). These solutions are as follows:
\begin{eqnarray}
\hskip-0.4cm
A &=& -p C, \quad B= \pm \sqrt{\mu-2 C^2}, \quad  \mu>2C^2,
\label{eq9}
\\
\hskip-0.4cm
A&=&-p C \pm \sqrt{\frac{\mu}{1+r}},~ B=\pm \sqrt{2 p A C},
~\mu>1+r.
\label{eq10}
\end{eqnarray}
and are simply the ``anti-phase-matched" and ``phase-matched" type solutions 
of Ref.~\cite{michal} respectively. Note that in addition to these solutions, 
there exists another one with no
free parameters (i.e., all amplitudes are directly dependent on $\mu$) which is a particular case of Eq. (\ref{eq10}),
namely
\begin{eqnarray}
A &=& p C, \quad B= \pm \sqrt{2 C^2}, \quad \mu={4 (1+r) C^2}.
\label{eq10a}
\end{eqnarray}

Let us now consider the GP Eqs.~(\ref{eq012})-(\ref{eq0})
looking for stationary solutions in that frame the GP equations become:
\begin{widetext}
\begin{eqnarray}
&& -is\partial_{x} \psi_{\pm 1} = -\frac{1}{2}\partial_{x}^{2} \psi_{\pm 1}
+ n_{\mathrm{tot}} \psi_{\pm 1} +
r (|\psi_{\pm 1}|^{2}+|\psi _{0}|^{2}-|\psi_{\mp 1}|^{2})\psi_{\pm 1}
+r \psi_{0}^{2}\psi_{\mp 1}^{\ast} - \mu \psi_{\pm 1},
\label{eq11}
\\
&& -is\partial_{x} \psi_{0} = - \frac{1}{2}\partial_{x}^{2} \psi_{0}
+n_{\mathrm{tot}} \psi_{0}+ r(|\psi_{-1}|^{2}+|\psi _{+1}|^{2})\psi _{0}
+2r \psi _{-1}\psi _{0}^{\ast }\psi_{+1} - \mu \psi_{0},
\label{eq12}
\end{eqnarray}
\end{widetext}
where we have slightly abused the notation by replacing the traveling wave variable
$\xi=x- s t$ with $x$ for simplicity.
We now decompose the amplitudes $R_j$ and phases $\phi_j$ of the order parameters
according to $\psi_j=R_j \exp(i \phi_j)$, and impose
the phase matching condition $\phi_1 + \phi_{-1}=2 \phi_0$,
to obtain the following equations:
\begin{eqnarray}
\partial_x \phi_1 &=& s \left(1 - (\frac{A}{R_1})^2\right),
\label{eq13}
\\[1.0ex]
\partial_x^2 R_1 &=& -s^2 \left(R_1 - \frac{A^4}{R_1^2}\right)
+2n_{\mathrm{tot}}R_1  \\
\notag&+&  2r (R_1^2+R_0^2-R_{-1}^2) R_1 +2 p r R_0^2 R_{-1}
-2\mu R_1,
\label{eq14}
\\[1.0ex]
\partial_x \phi_0 &=& s \left(1 - (\frac{B}{R_0})^2\right),
\label{eq15}
\\[1.0ex]
\partial_x^2 R_0 &=& -s^2 \left(R_0 - \frac{B^4}{R_0^2}\right)
+ 2n_{\mathrm{tot}}R_0 \\
\notag&+& 2 r (R_1^2+R_{-1}^2) R_0 + 4 p r R_0 R_{1} R_{-1}
-2\mu R_0,
\label{eq16}
\\[1.0ex]
\partial_x \phi_{-1} &=& s \left(1 - (\frac{C}{R_{-1}})^2\right),
\label{eq17}
\\[1.0ex]
\partial_x^2 R_{-1} &=& -s^2 \left(R_{-1} - \frac{C^4}{R_{-1}^2}\right)
+ 2n_{\mathrm{tot}}R_{-1} \\
\notag&+& 2 r (R_{-1}^2+R_0^2-R_{1}^2) R_{-1} +2 p r R_0^2 R_{1}
-2\mu R_{-1}.
\label{eq18}
\end{eqnarray}
Notice that in these equations, the asymptotic states $A$, $B$, and $C$
arise naturally due to the integration of the equations for the phases.

In order to seek instabilities of the steady state flow at different
fluid speeds, we now linearize around the asymptotic states, according to
$R_1=A + \epsilon r_1(x)$, $R_0=B + \epsilon r_0(x)$ and
$R_{-1}=C + \epsilon r_{-1}(x)$ (where $\epsilon$ is a formal small parameter).
Substituting the above expressions
into Eqs.~(\ref{eq13})--(\ref{eq18}), we obtain a system of three
second-order ordinary differential equations;
the latter, can be
readily expressed as a system of six first-order equations of the following form:
\begin{equation}
\notag
\frac{d}{dx}
\left(\begin{array}{c}
r_1 \\[0.5ex]
\displaystyle
       r_1'     \\[0.5ex]
r_2 \\ [0.5ex]
\displaystyle
       r_2'     \\[0.5ex]
r_3 \\ [0.5ex]
\displaystyle
       r_3'
\end{array} \right)
= M
\left(\begin{array}{c}
r_1 \\ [0.5ex]
\displaystyle
       r_1'     \\[0.5ex]
r_2 \\ [0.5ex]
\displaystyle
       r_2'     \\[0.5ex]
r_3 \\ [0.5ex]
\displaystyle
       r_3'
\end{array} \right),
\end{equation}
where $r_j'\equiv dr_j/dx$ and
\begin{equation}
\notag
M \equiv \{m_{ij}\} =
\left(
\begin{array}{cccccc}
0& 1  & 0 & 0 & 0 & 0 \\[0.7ex]
m_{21} & 0 & m_{23} & 0 & m_{25} & 0 \\[0.7ex]
0 & 0 & 0 & 1 & 0 & 0 \\[0.7ex]
m_{41} & 0 & m_{43} & 0 & m_{45} & 0 \\[0.7ex]
0 & 0 & 0 & 0 & 0 & 1 \\[0.7ex]
m_{61} & 0 & m_{63} & 0 & m_{65} & 0
\end{array}
\right),
\end{equation}
and the non-zero matrix elements of $M$ are
given by the following expressions:
\begin{eqnarray}
\notag
m_{21}  &=&  -4 s^2 +2 (3A^2+B^2+C^2) +\\
\notag
\notag &&  2 r (3 A^2+B^2-C^2)-2 \mu,
%\label{eq19}
\\[1.0ex]
\notag
m_{23} &=&m_{41}= 4 A B (1+r) + 4 p r B C,
%\label{eq20}
\\[1.0ex]
\notag
m_{25} &=&m_{61}= 4 A C (1-r) +2 p r B^2,
%\label{eq21}
\\[1.0ex]
\notag
m_{43} &=&  -4 s^2 +2 (3B^2+A^2+C^2) +\\
\notag
\notag &&  2 r (A^2+C^2) + 4 p r A C -2 \mu,
%\label{eq23}
\\[1.0ex]
\notag
m_{45} &=&m_{63}= 4 B C (1+r) +4 p r A B,
%\label{eq24}
\\[1.0ex]
\notag
m_{65} &=&  -4 s^2 + 2 (3 C^2+B^2+A^2) +\\
\notag
\notag &&  2 r (3 C^2+B^2-A^2)-2 \mu.
%\label{eq27}
\end{eqnarray}
Notice that in order to derive the above system of ordinary differential equations,
we have partially simplified the problem, assuming
no perturbations in the phases. In such a more general case, however,
the full first-order ODE system incorporating phase perturbations
is in fact twelve-dimensional and is not analytically tractable.
We have found (not treated explicitly here) that this system can be analyzed
only in some special cases, such as $B=0$,
yielding the same results for the critical defect speeds, as will be
presented below (see also the discussion of Section II.B in Ref.~\cite{ourbd}).

We now follow the approach used for the one-component GP equation
in Ref. \cite{hakim} (subsequently generalized in the two-component case in Ref. \cite{susanto})
to determine the critical speeds of the defect. In particular,
the critical speeds can readily be found upon computing
the eigenvalues of the matrix $M$
and setting them equal to zero.
The violation of this 
threshold condition is
tantamount to the emergence of a number of oscillatory modes that  
enforce too many constraints and prevent the existence of localized
solutions for a generic obstacle potential, as is explained in detail in
Ref. \cite{hakim}.
It is straightforward to examine this condition 
both in the case of $p=1$ and of $p=-1$.
We will demonstrate below the case of $p=1$ for definiteness.
In this case, by considering the stationary state of the form of
Eq.~(\ref{eq9}), we obtain two different speeds of sound, namely,
\begin{eqnarray}
c_1
=\sqrt{r}c_2, \quad c_2=\sqrt{\mu}.
\label{eq28}
\end{eqnarray}
It is clear that the first critical velocity is characteristic
for the spinor $F=1$ condensate under consideration (as it depends on
the normalized spin-dependent interaction strength $r$), while the
second one is the standard speed of sound appearing in the one-component
GP equation~\cite{hakim} (note that this speed may also appear in the case of the
spinor $F=1$ BEC as well \cite{ourbd}).
It is worthwhile to point out that as $r>0$
for {\it anti-ferromagnetic} bosonic spin-1 atoms (e.g., $^{23}$Na), while
it is $r<0$ for {\it ferromagnetic} ones (e.g., $^{87}$Rb), then
the first speed of sound is relevant (i.e., will only exist)
in the case of, e.g., the polar $^{23}$Na spinor condensate.
On the other hand, by selecting the stationary states of the form
of Eq.~(\ref{eq10}), then again we find two critical speeds, which are now given by:
\begin{eqnarray}
c_1
=\sqrt{\frac{-r}{1+r}}c_2,
\quad c_2=\sqrt{\mu}.
\label{eq29}
\end{eqnarray}
In this case, it is clear that the first critical speed
will exist only in the ferromagnetic spinor BECs (such as $^{87}$Rb),
but not in anti-ferromagnetic ones (such as $^{23}$Na); nevertheless,
it should be noted that since the normalized spin-dependent interaction
strength is small in both cases of $^{87}$Rb and $^{23}$Na condensates
($r=O(10^{-2})$ as discussed above), the lower critical speeds are
approximately the same.

We now test these analytical predictions by dragging a localized defect
(e.g., a blue-detuned laser beam)
through the condensate at different speeds characterizing the three
regimes, namely (a) $0<s<c_1<c_2$, (b) $c_1<s<c_2$, and (c) $c_1<c_2<s$.

\begin{figure}[!ht]
\includegraphics[width=8.5cm]{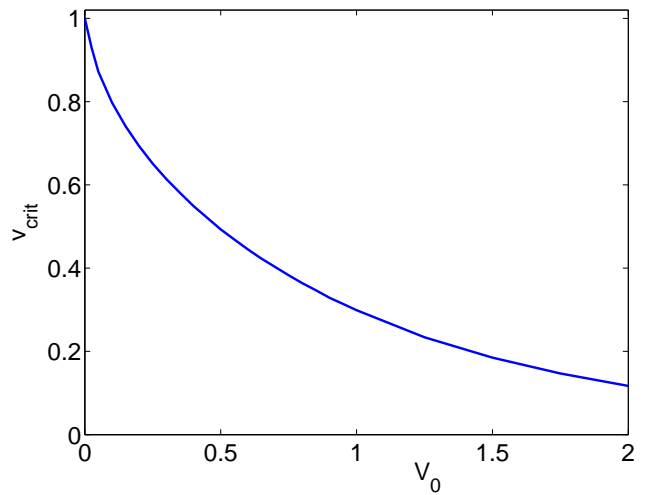}
\caption{Dependence of the critical velocity on the defect
strength $V_0$. The result shown
corresponds to the stationary solution of Eq.~(\ref{eq9})
with $A=-0.5=-C$ and $B=\sqrt{1/2}$ ($\mu=1$).
It was confirmed that, e.g., for the solutions of
Eq.~(\ref{eq10}), the analytical and numerical results were indistinguishable up
to three decimal places.
}
\label{fig1}
\end{figure}

%c_2=0.329$ and $c_1=\sqrt(r)*c_2= 0.177*0.329=0.058$

\begin{figure}[!ht]
\includegraphics[width=8.5cm]{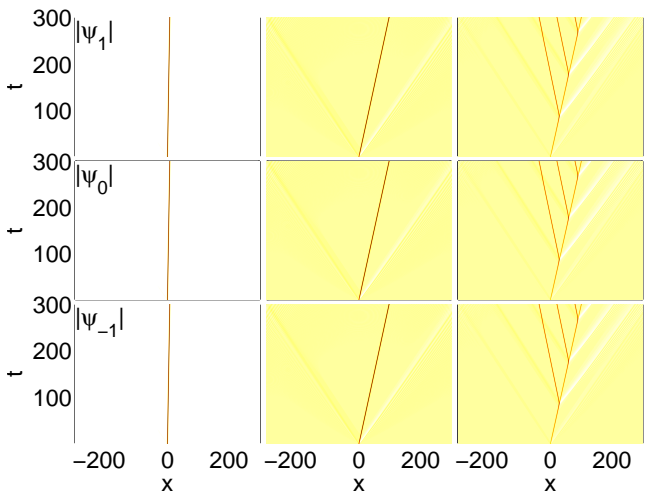}
\caption{(Color online) Time evolution of the three hyperfine components
(top, middle and bottom rows, respectively) for different defect velocities:
$s=0.025 < c_1$ (left column),
$c_1<s=0.325 \lesssim c_2$ (middle column)
and
$s=0.335 > c_2$ (right column);
here, $c_1=0.058$ and $c_2 = 0.329$ are the two
critical velocities, while
the defect strength is taken to be $V_0=0.9$.
The initial condition corresponds to the
stationary state with wavefunction amplitudes given in Eq.~(\ref{eq9}).
While the results in the left and middle columns show a steady flow
(apart from an oscillatory structure that is detached at $t=0$)
the evolution shown in the panels of the right column is characterized by
the emission of dark solitons, even from the early stages of the process.
Notice that the analytically predicted first (lower)
critical velocity would fall between the velocities of the results
depicted in the left and middle columns but no significant change
is observed in the dynamics between these two cases.
}
\label{figBdyn}
\end{figure}

\begin{figure}[!ht]
\includegraphics[width=8.5cm]{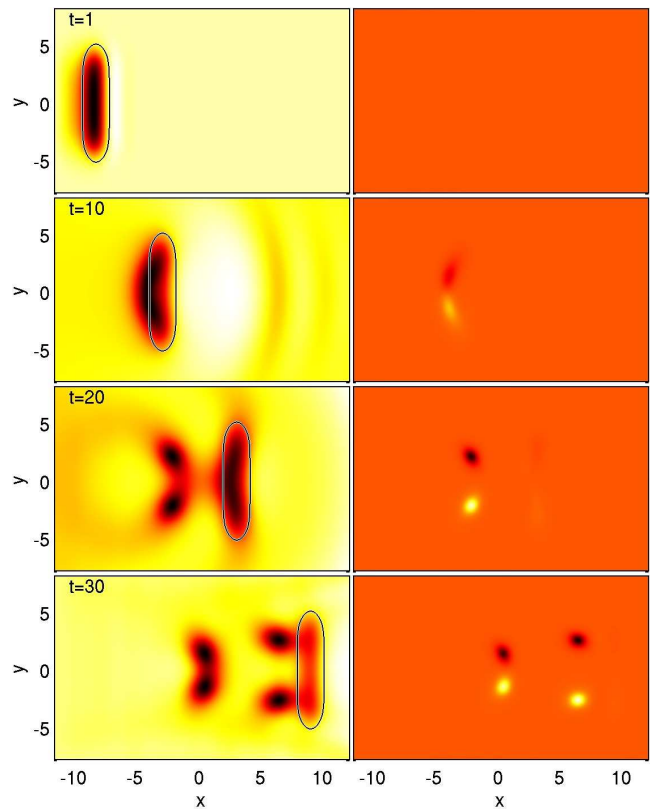}
\caption{(Color online)
Vortex and anti-vortex pairs nucleated by a moving defect in the
2D spinor condensate.
The left column shows the density of the
spinor condensate's component $\psi_0$ at different times (indicated
in the panels).
The location and extent of the moving defect is depicted by the
oval line corresponding to an iso-contour of its strength
at 10\% of its maximum.
The right column shows the
corresponding vorticity (defined in the text), clearly
illustrating the presence of vortex and anti-vortex pairs.
The case depicted here corresponds to $w_y=8$, $a=2$,
and $V_0=0.9$ while the defect speed is taken to be $s=0.6>c_2$.}
\label{fig2Dw8}
\end{figure}

\begin{figure}[!ht]
\begin{center}
\includegraphics[height=6.75cm]{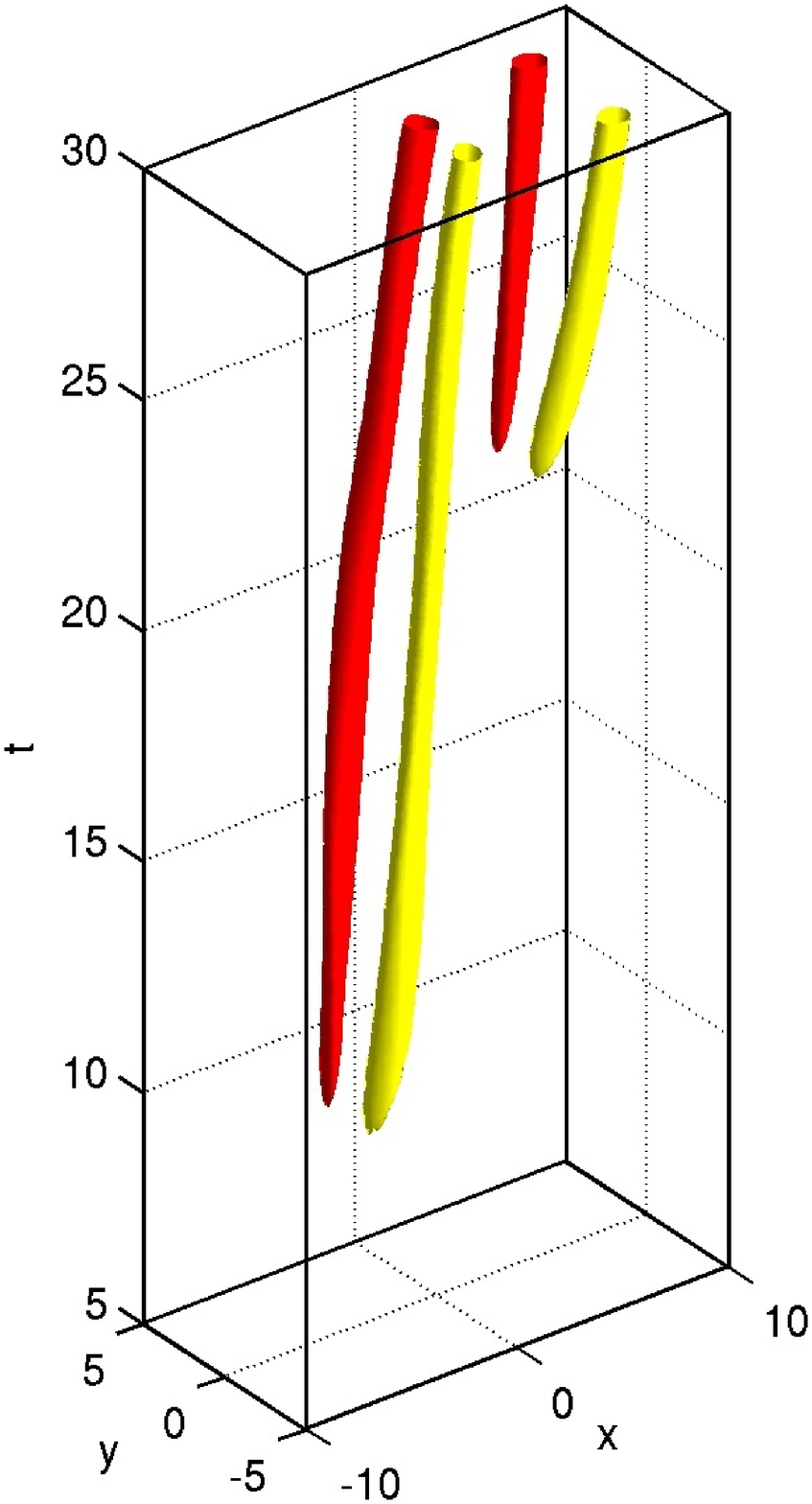}
\includegraphics[height=6.75cm]{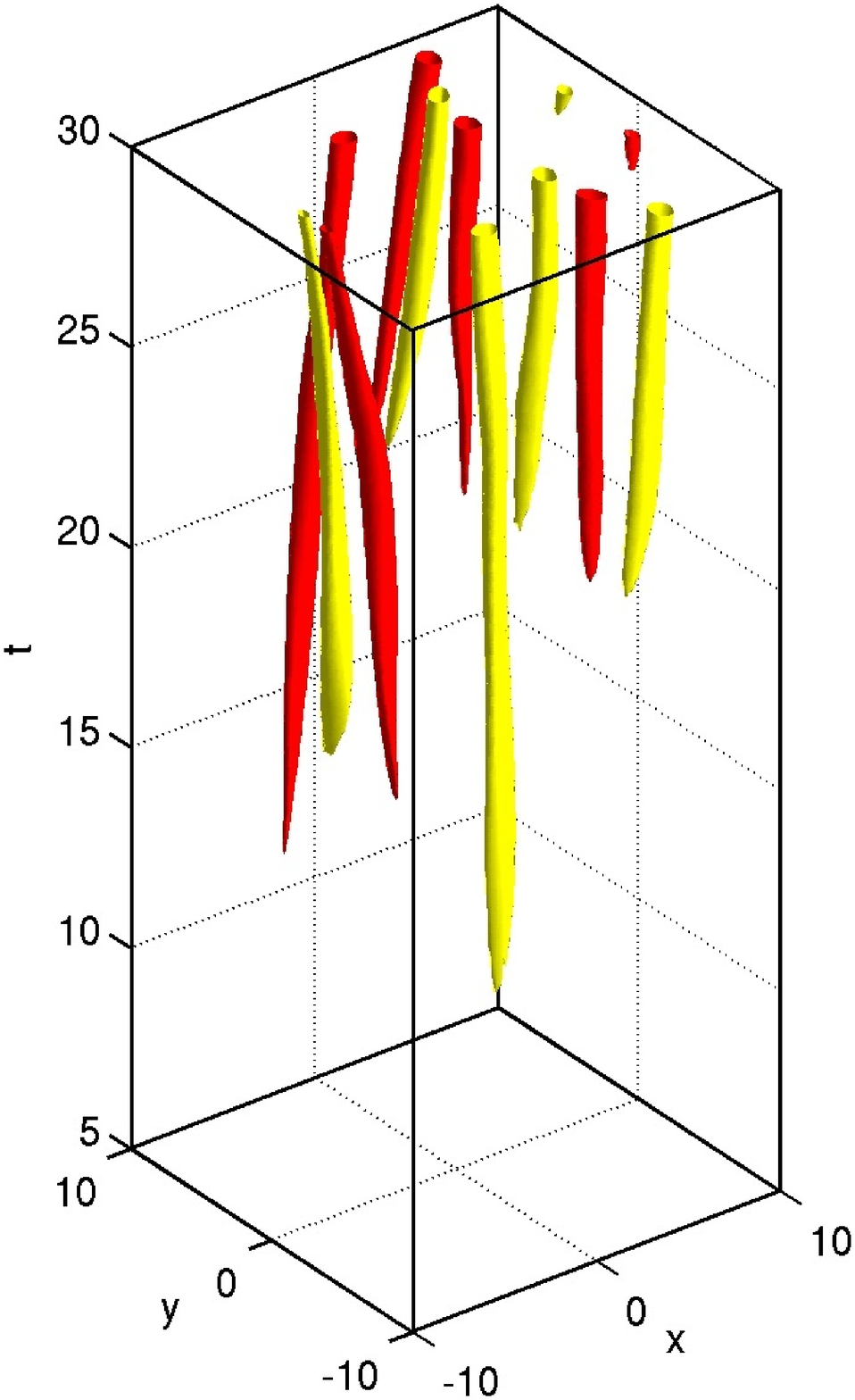}
\end{center}
\caption{(Color online) Evolution of the maximum (dark/red) and
minimum (light/yellow) of the vorticity iso-contours in $(x,y,t)$.
One can clearly discern the emergence of vortex (in dark/red) and
antivortex (in light/yellow) pairs, as the defect moves along the
$x$-direction.
The left and right panels correspond, respectively, to the
cases depicted in Figs.~\ref{fig2Dw8} ($w_y=8$) and \ref{fig2Dw20} ($w_y=20$).
}
\label{fig2Disow8}
\end{figure}

\begin{figure}[!ht]
\includegraphics[width=8.5cm]{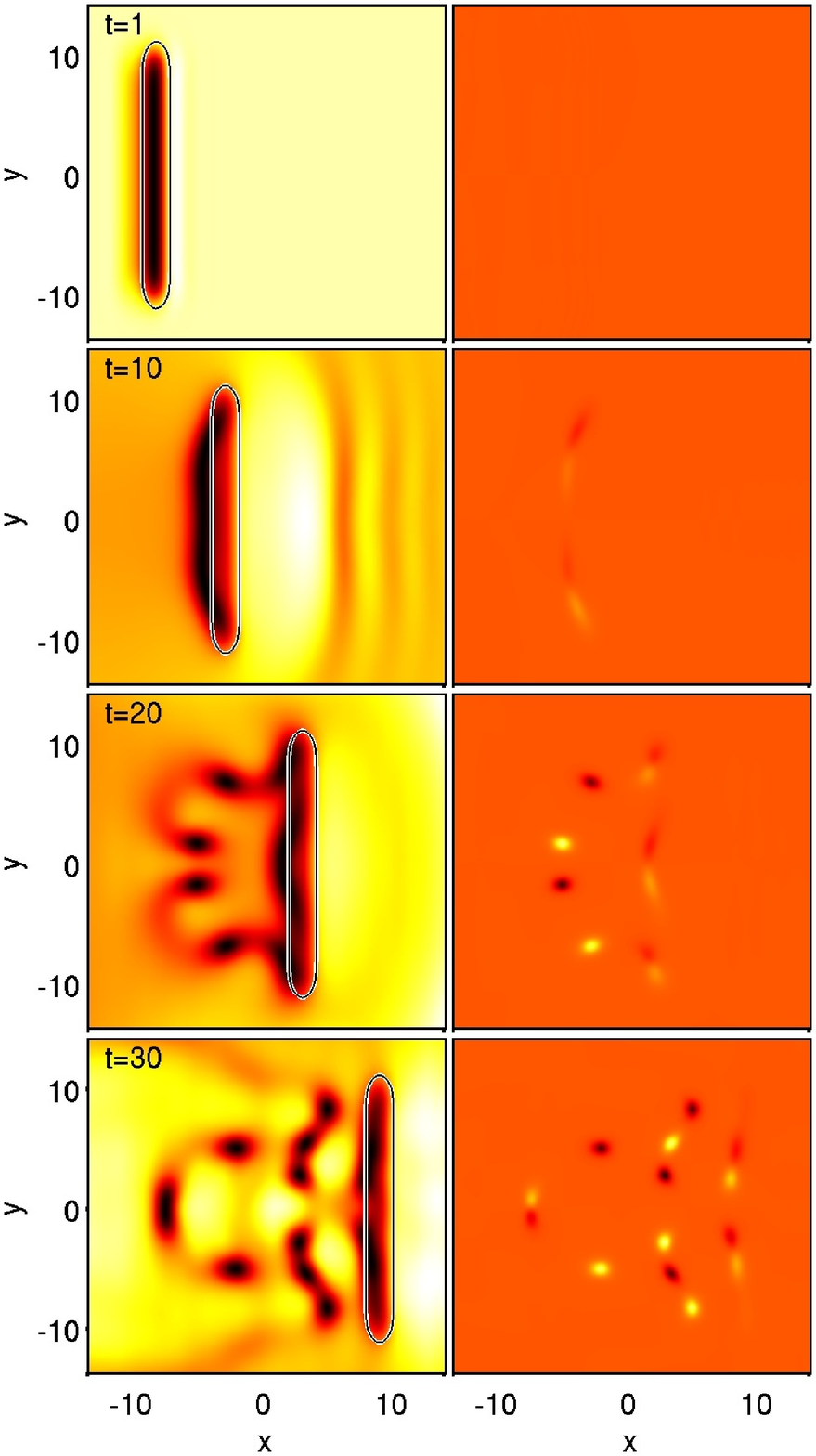}
\caption{(Color online) Same as Fig. \ref{fig2Dw8} but for a wider
defect with $w_y=20$. One can clearly observe the formation of numerous
vortex anti-vortex pairs in the wake of the defect.
}
\label{fig2Dw20}
\end{figure}

\section{Results of Numerical Studies}

\subsection{One-dimensional setting}

Our more detailed results concern the 1D setting,
where we explore the full two-parameter space of speeds $s$
and defect strengths $V_0$, for $a=2$ in the case of the anti-ferromagnetic
$^{23}$Na spinor BEC, characterized by the spin-dependent interaction strength
$r=0.0314$.
Figure \ref{fig1} illustrates the threshold above which coherent
localized excitations are emitted from the defect as it
propagates through the condensate. A typical example of the
evolution process for
(a) $0<s=0.025<c_1<c_2$, (b) $c_1<s=0.325<c_2$ and (c) $c_1<c_2<s=0.335$
is shown in Fig. \ref{figBdyn}. Several comments are in order here:
\begin{itemize}
\item As expected from the analytical predictions, when the defect speed is
below both critical values $c_1=\sqrt{\mu r}$, and  $c_2=\sqrt{\mu}$, the defect
moves through the atomic cloud without emission of any nonlinear excitation.
An oscillatory structure is
radiated at the initial time (similarly to what has been observed earlier,
e.g., in Ref. \cite{susanto}) both at the front,
as well as at the
rear of the defect, moving with the speed of sound; however,
no further such radiation is observed.
\item Remarkably, for speeds intermediate between $c_1$ and $c_2$,
we do {\it not} observe any modification in the dynamics.
This means that the first critical speed $c_1$
does not appear to be activated by the system.
This finding is even more surprising in light of the fact
that for $r < 0$, this critical speed has been recognized to be
directly connected to the quasi-momentum (wavenumber) associated with the
modulational instability of the ferromagnetic spinor
condensate \cite{ostro} (see also the relevant discussion in Ref. \cite{ourbd}).
Nevertheless, in all of our simulations, both in 1D and in higher
dimensions, we have definitively confirmed the apparent physical
irrelevance of this first
critical speed (which is the lower nontrivial critical speed in the spinor BEC case).
It should be noted here that this same feature has been confirmed
for cases where the spin-dependent interaction strength $r$
was artificially increased to considerably larger values (by an
order of magnitude in comparison with its physically relevant value of
$r=3.14 \times 10^{-2}$ for $^{23}$Na).
\item
When the defect speed is larger than both critical ones, there is a
clear emission of dark (in fact, gray) solitons,
which travel in a direction opposite to that defect, with velocities less
than the speed of sound. Similarly to one-
\cite{hakim,radouani,george,our1d} and two-component \cite{susanto} settings,
the solitons temporarily ``alleviate'' the super-critical nature of
the flow, but eventually they are separated enough from the defect that
another such excitation emerges. For this reason, the emission seems to
be regularly spaced as shown in the right panels of Fig. \ref{figBdyn}.
\item As the strength of the defect $V_0 \rightarrow 0$,
the critical speed $c_2$ observed from the numerical simulations
tends asymptotically to the one theoretically
predicted from the
analysis above, i.e., $\sqrt{\mu}=1$ in this case (this feature
has been confirmed for different values of $\mu$, such 
as $\mu=2$ and $\mu=4$). However,
similarly to what was observed in Refs.~\cite{hakim,susanto}, as the
strength of the defect increases, the value of the critical
speed accordingly decreases (since nucleation of dark solitons
is easier for the lower density BEC).
\item Finally, we note that in all our simulations (even in higher dimensions,
see below) the three spinor components were locked to each other through
$|\psi_1|^2/A^2 \approx |\psi_0|^2/B^2 \approx |\psi_{-1}|^2/C^2$. This
tight restriction is presumably related to the fact that we only
observed one critical nucleation speed in our simulations: as
all the components are tightly locked to each other, they behave
like a single component and thus only one critical speed is
observed.
\end{itemize}

\subsection{Two-dimensional setting}

In the
2D case, motivated by the recent work of Ref. \cite{engels} in the single
component case, 
we consider a defect which is localized
along the $x$-axis (with a width $a^{-1}$) but elongated along the $y$-axis
(with a width $w_y > a^{-1}$), namely:
\begin{eqnarray}
V&=&\frac{V_0}{4}\, \exp\left(-a (x-s t)^2\right)
%\label{sub2d}
\notag
\\&\times&
\notag
\left[\tanh\left(y+\frac{w_y}{2}\right)+1\right]
\left[\tanh\left(-y+\frac{w_y}{2}\right)+1\right].
\end{eqnarray}
Once again, we find that (i) no emission of nonlinear excitations is present
for $s<c_2$ and that (ii) the emission of nonlinear excitations,
which now have the form of vortex-antivortex pairs, arises for speeds
larger than the critical speed $c_2$. Figure \ref{fig2Dw8} illustrates
the case of $s=0.6>c_2$ for the defect strength $V_0=0.9$, while
the defect width along the $y$-direction is $w_y=8$.
In addition to showing the density, the figure shows
the vorticity defined as $\omega=\nabla \times v_{\rm f}$, where
the fluid velocity $v_{\rm f}$ is given by
\begin{eqnarray}
v_{\rm f}= \frac{\psi^{\star} \nabla \psi- \psi \nabla \psi^{\star}}{i |\psi|^2},
\label{veloc}
\end{eqnarray}
for a given
hyperfine-component $\psi$. In the contour plots of the
vorticity $\omega$, the emergence of vortex-antivortex pairs
is immediately evident in the supercritical case shown.
In fact, in order to provide a more clear sense of the temporal
dynamics and the nucleation of the coherent structure pairs,
we show in the left panel of Fig.~\ref{fig2Disow8}
the spatiotemporal evolution of the
iso-contours of vorticity for the same numerical simulation as in
Fig.~\ref{fig2Dw8}. The vorticity renders transparent the emergence
of the different vortex pairs at different moments in time
(and accordingly different locations in $x$, as the defect travels).

Finally, in Fig. \ref{fig2Dw20}, we also show a case example of a
considerably wider defect, with a width $w_y=20$.
It can be seen that in this setting, the region of low density caused by
the defect is far wider, in turn leading to a breakup
into a large number of vortex pairs that can
be identified not only by the density minima, but also
even more clearly (including their topological charge)
by the vorticity panels. In the right panel of
Fig.~\ref{fig2Disow8} we depict the corresponding spatiotemporal
evolution of the vorticity.
We note that, as it was the case in the 1D setting, all our
simulations suggest that the three spinor components remain
essentially locked satisfying the relation $|\psi_1|^2/A^2 \approx
|\psi_0|^2/B^2 \approx |\psi_{-1}|^2/C^2$.
For this reason, we only depict the dynamics of the $\psi_0$
component in all of our results.

\begin{figure*}[!ht]
\begin{center}
\includegraphics[width=4.5cm]{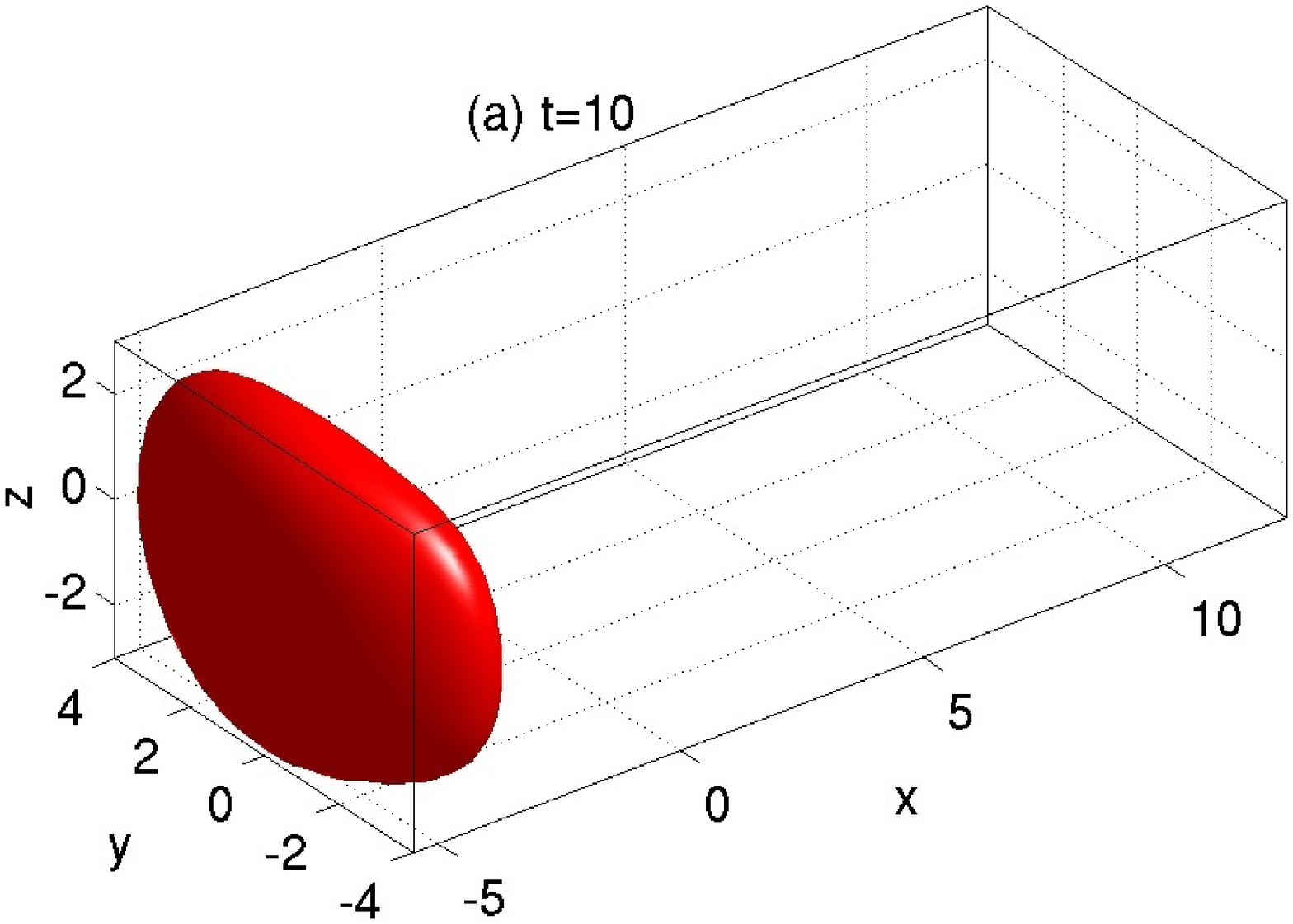}~~~~
\includegraphics[width=4.5cm]{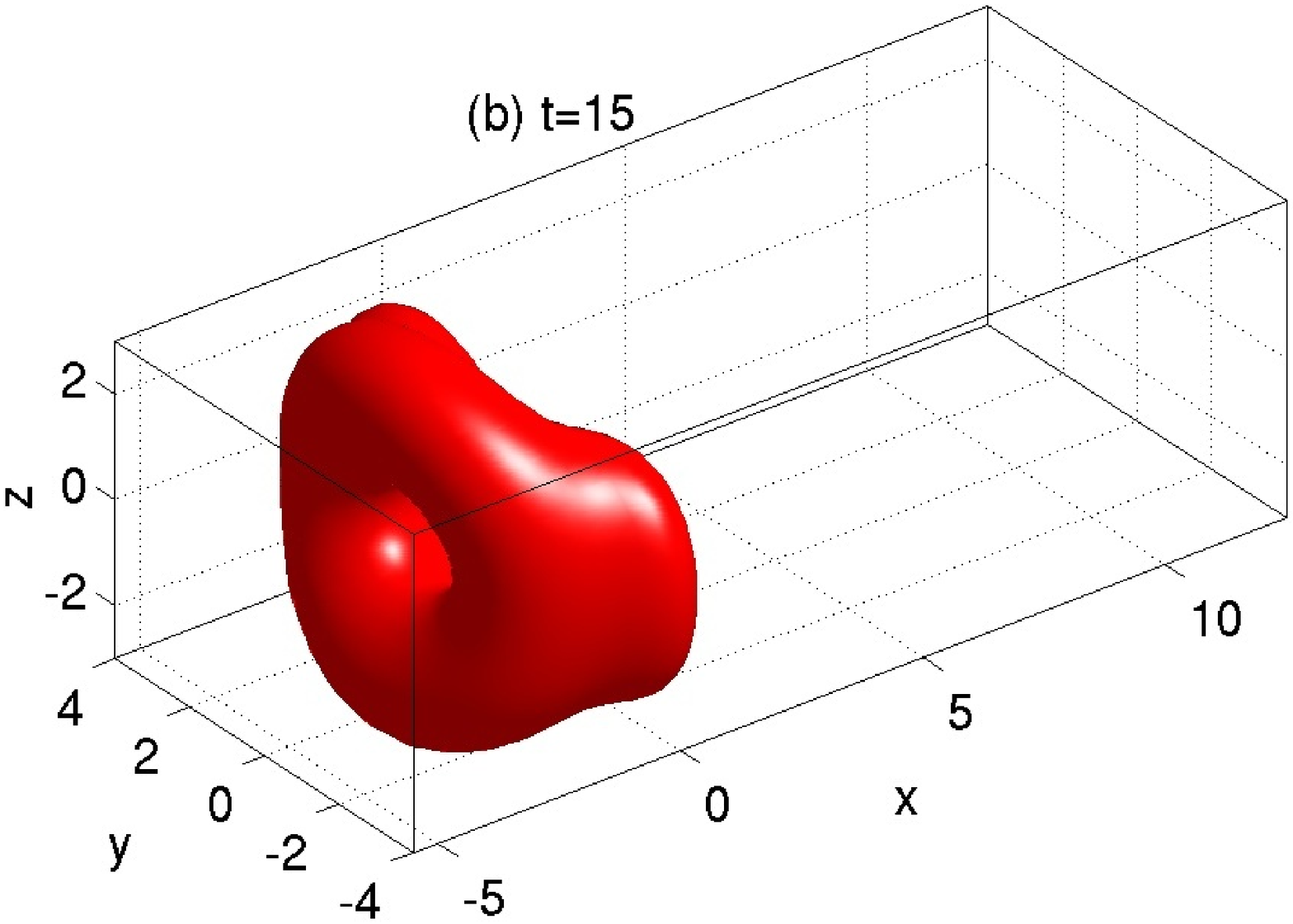}~~~~
\includegraphics[width=4.5cm]{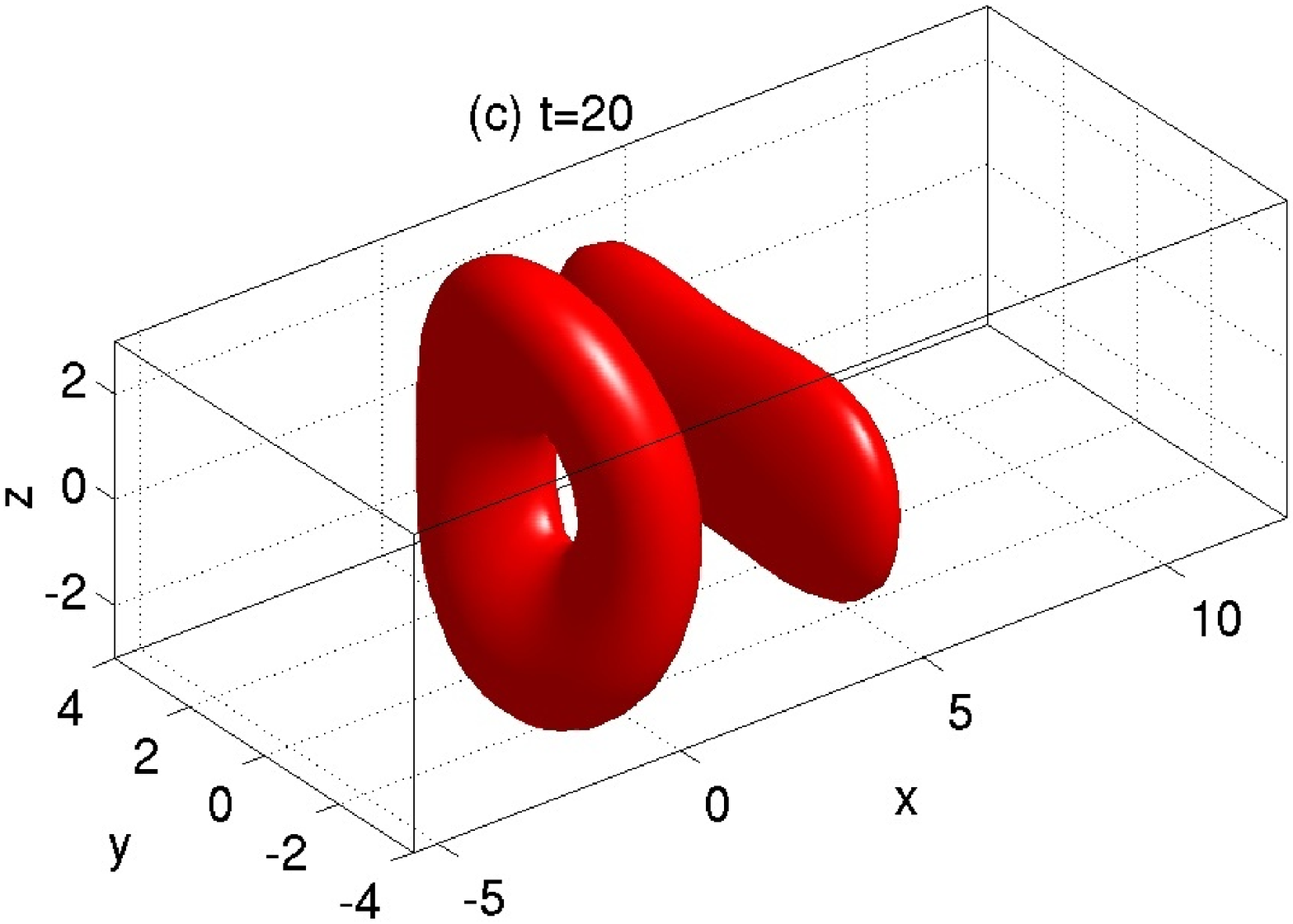}\\
\includegraphics[width=4.5cm]{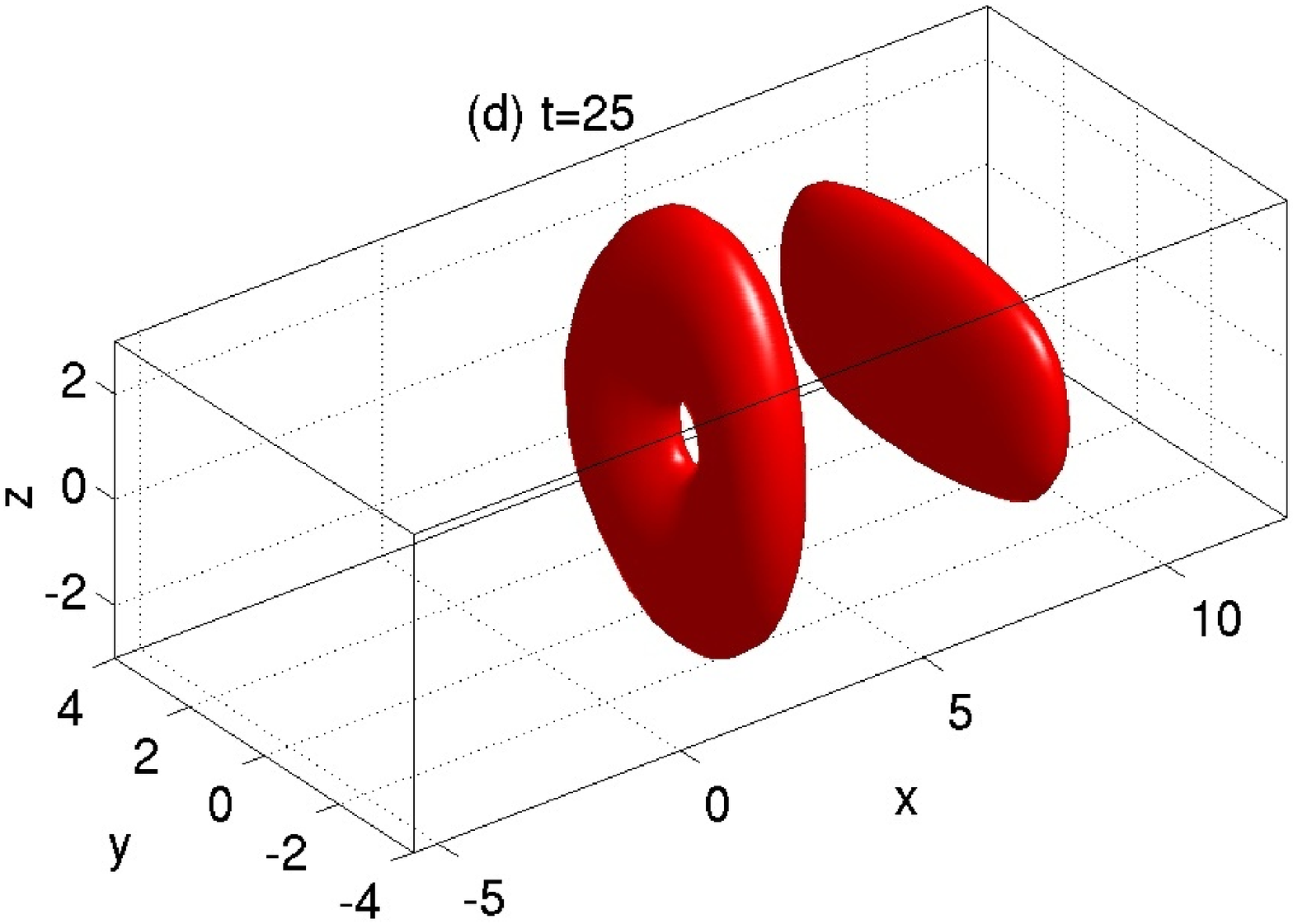}~~~~
\includegraphics[width=4.5cm]{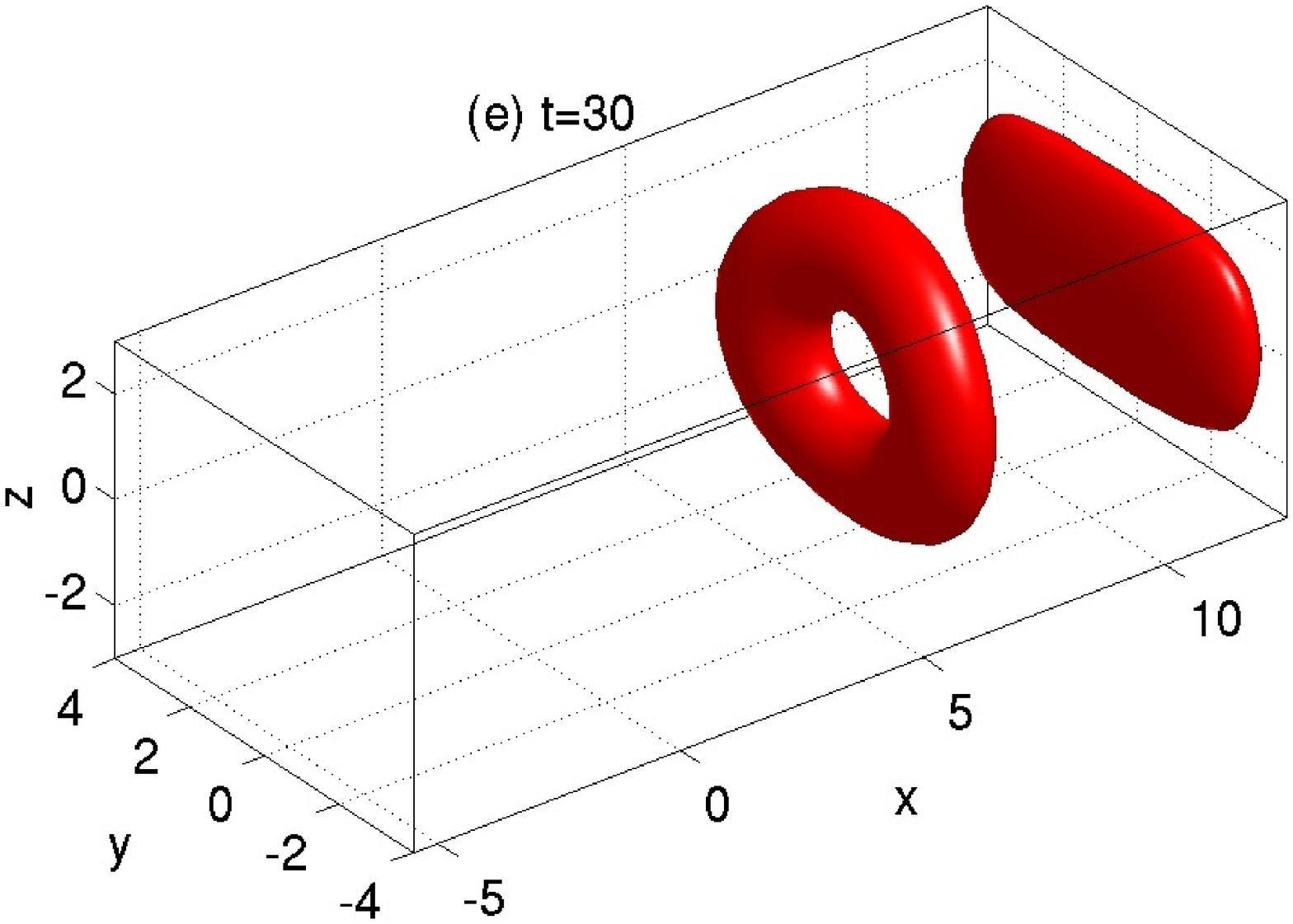}~~~~
\includegraphics[width=4.5cm]{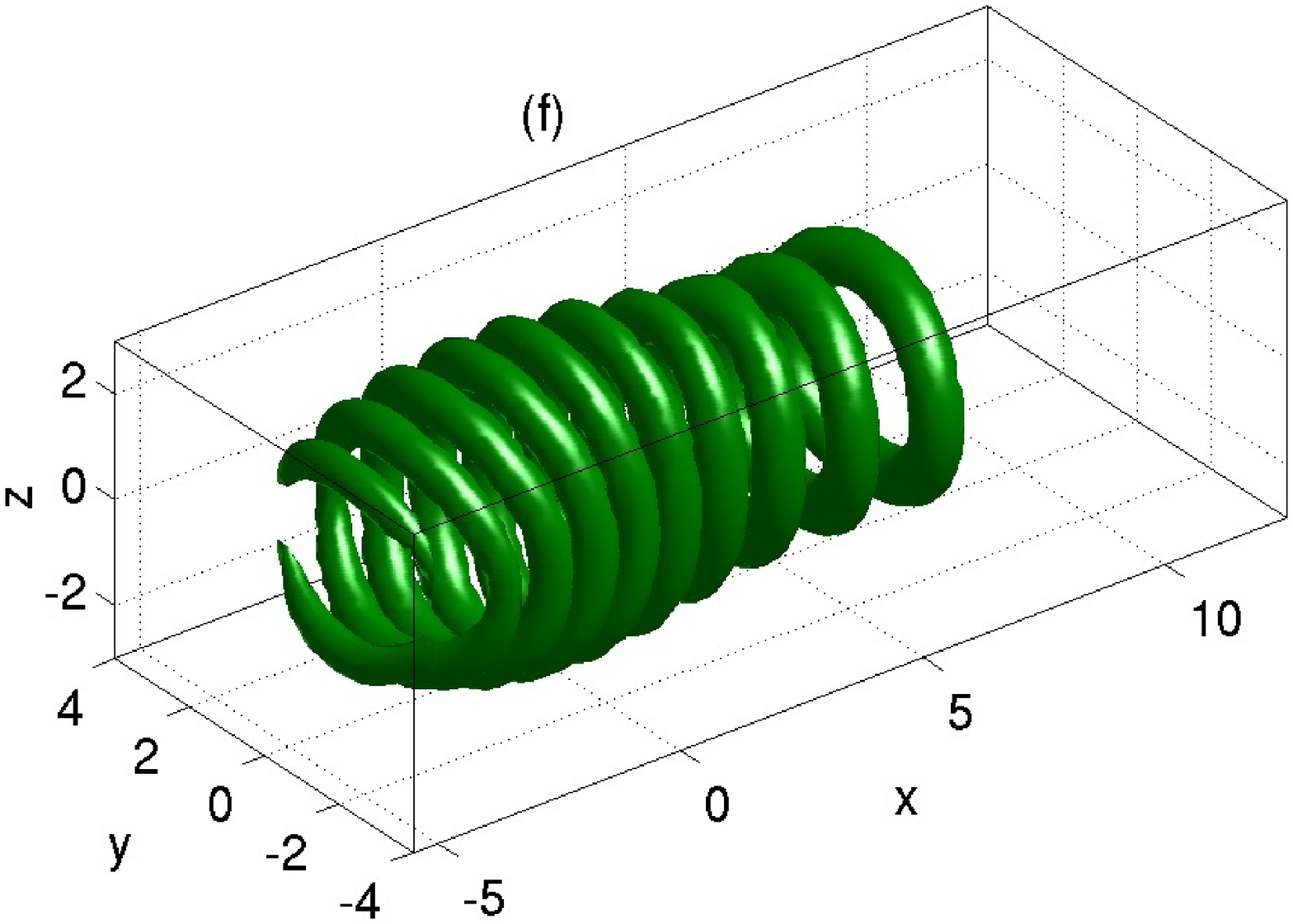}
\end{center}
\caption{(Color online)
Vortex ring formation in the supercritical 3D setting.
Panels (a)--(e) depict iso-density contours corresponding to
$|\psi_0|^2=0.3$ at the indicated times. Note that the
extent of the moving defect is clearly visible in these panels
(it corresponds to the rightmost flat oval shape that is
created by the atomic density depletion due to its presence).
Panel (f) shows typical isocontours of the norm of the vorticity
of $\psi_0$ at times $t=12,14,\dots,30$ (left to right).
In this case we use a defect with $w_y=8$, $w_z=4$, $a=2$ and
speed $s=0.8>c_2$.
}
\label{fig3Dw8vortvel08}
\end{figure*}

\begin{figure*}[!ht]
\begin{center}
\includegraphics[width=2.90cm]{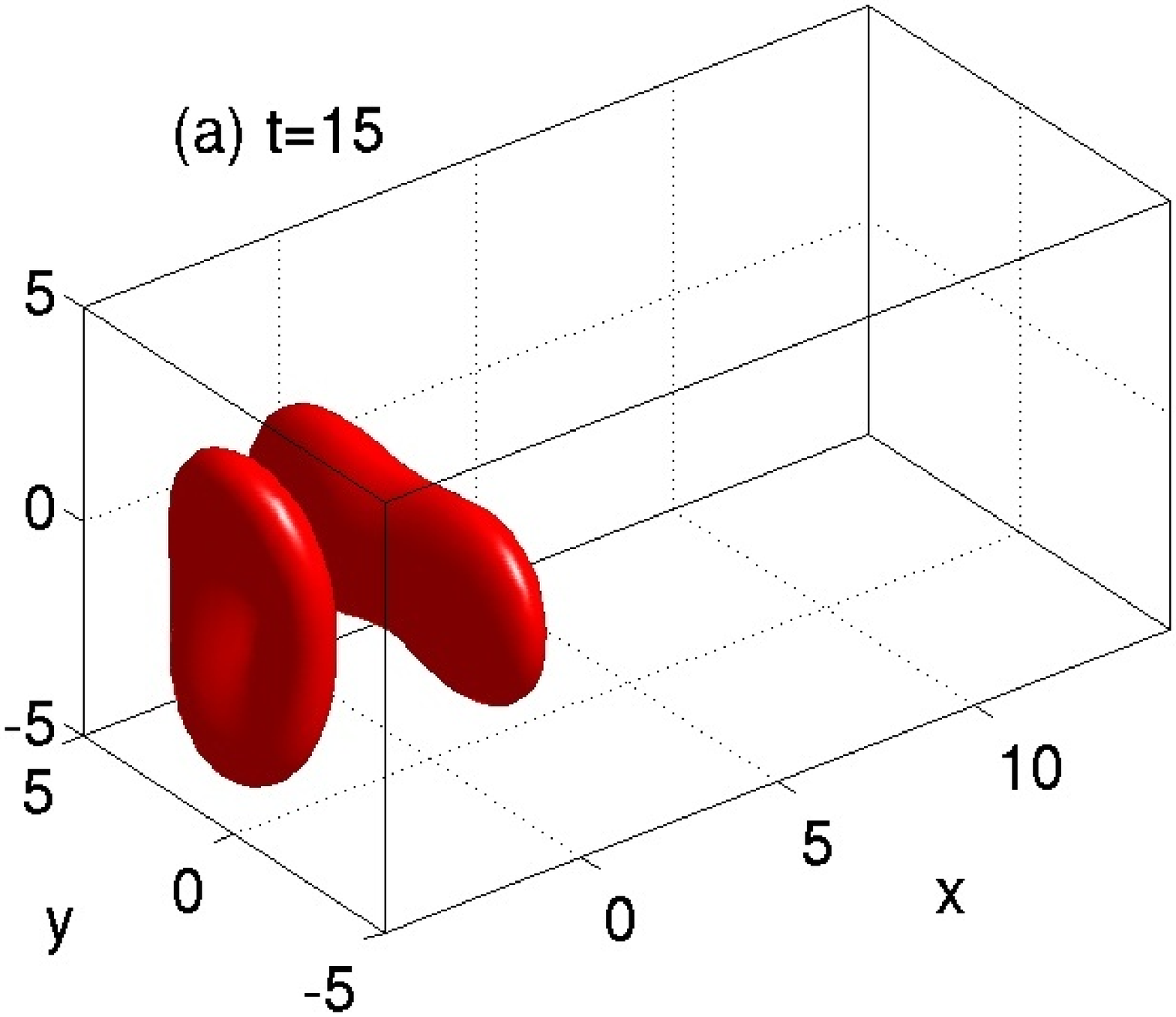}
\includegraphics[width=2.90cm]{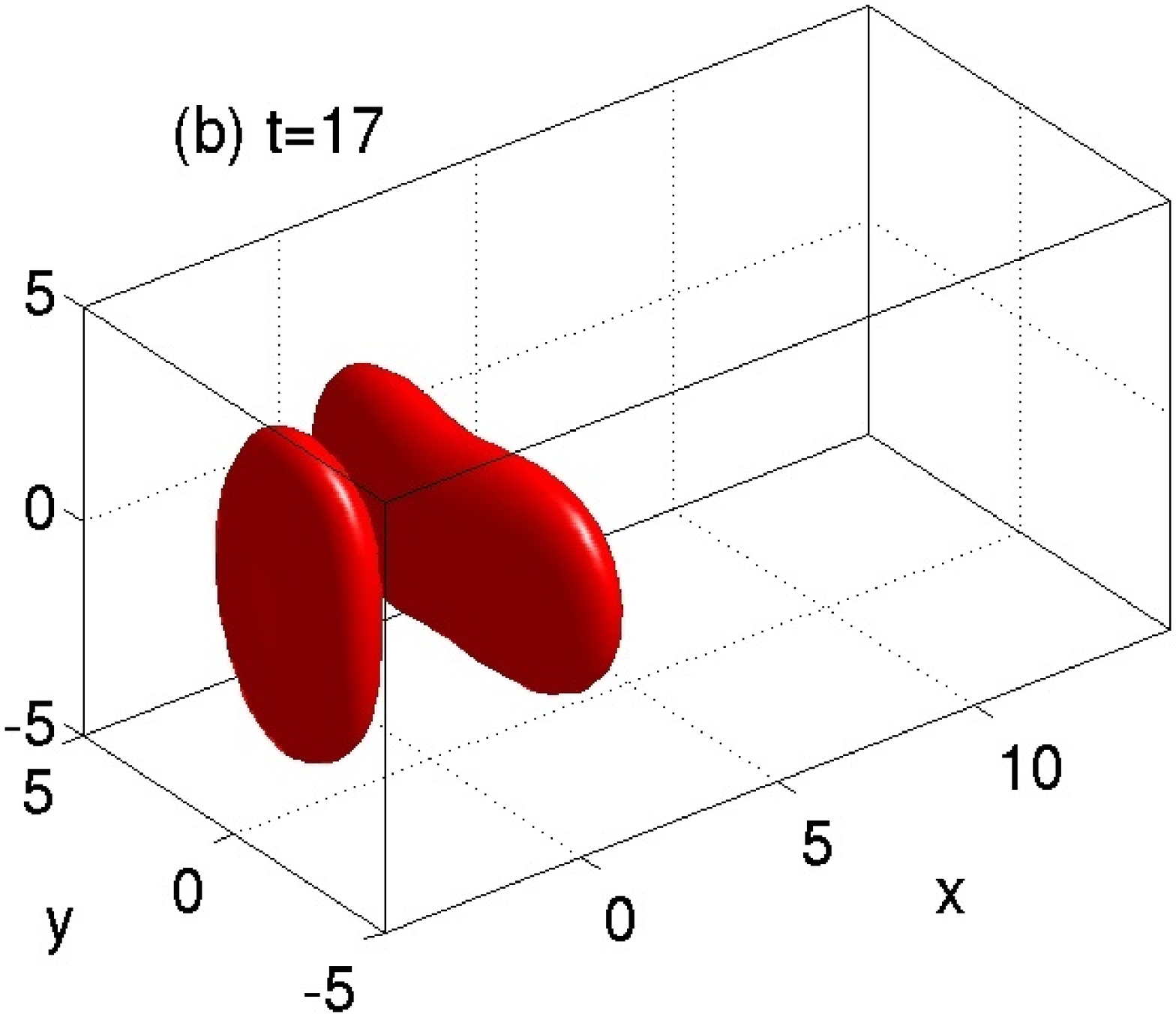}
\includegraphics[width=2.90cm]{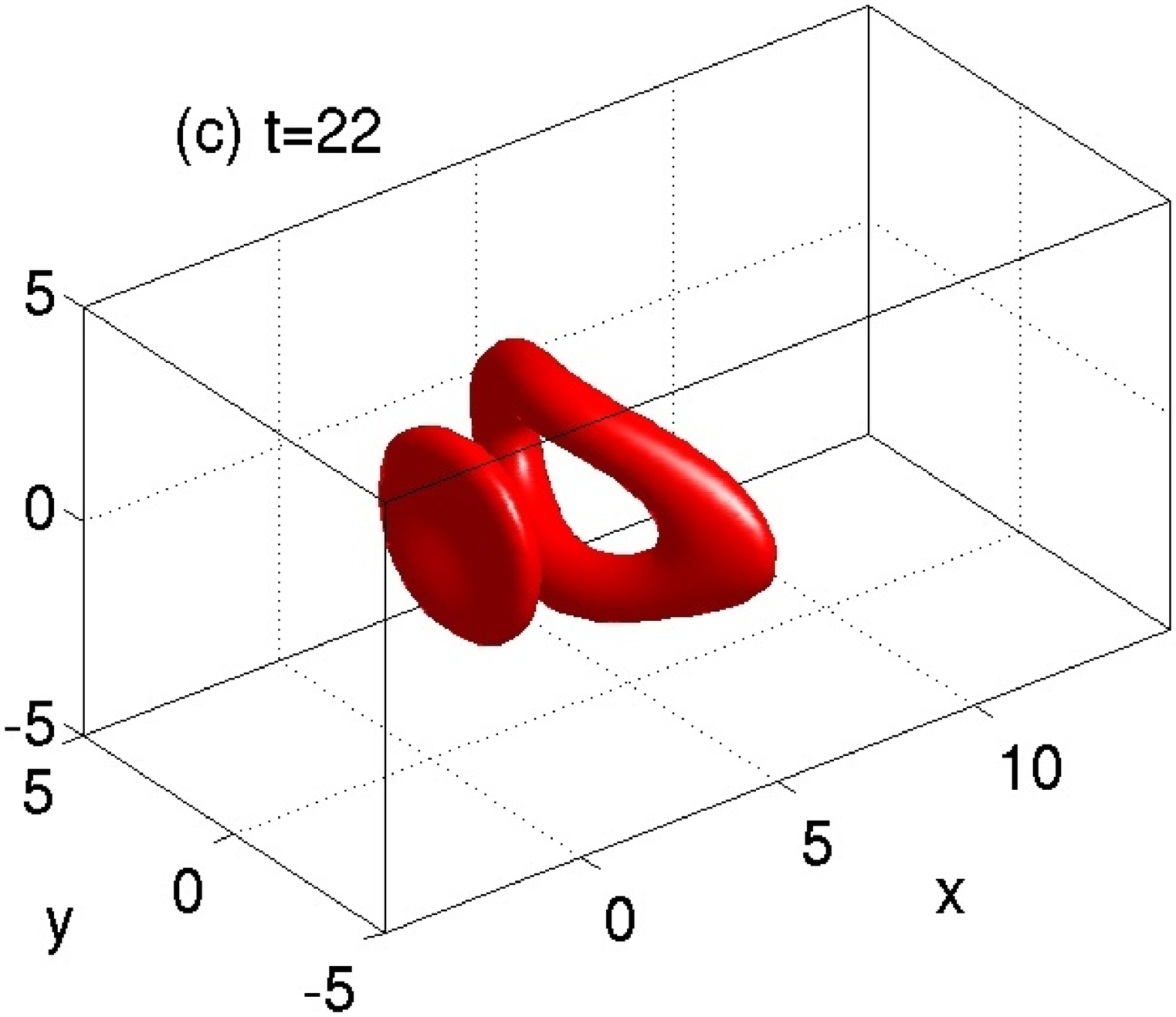}
\includegraphics[width=2.90cm]{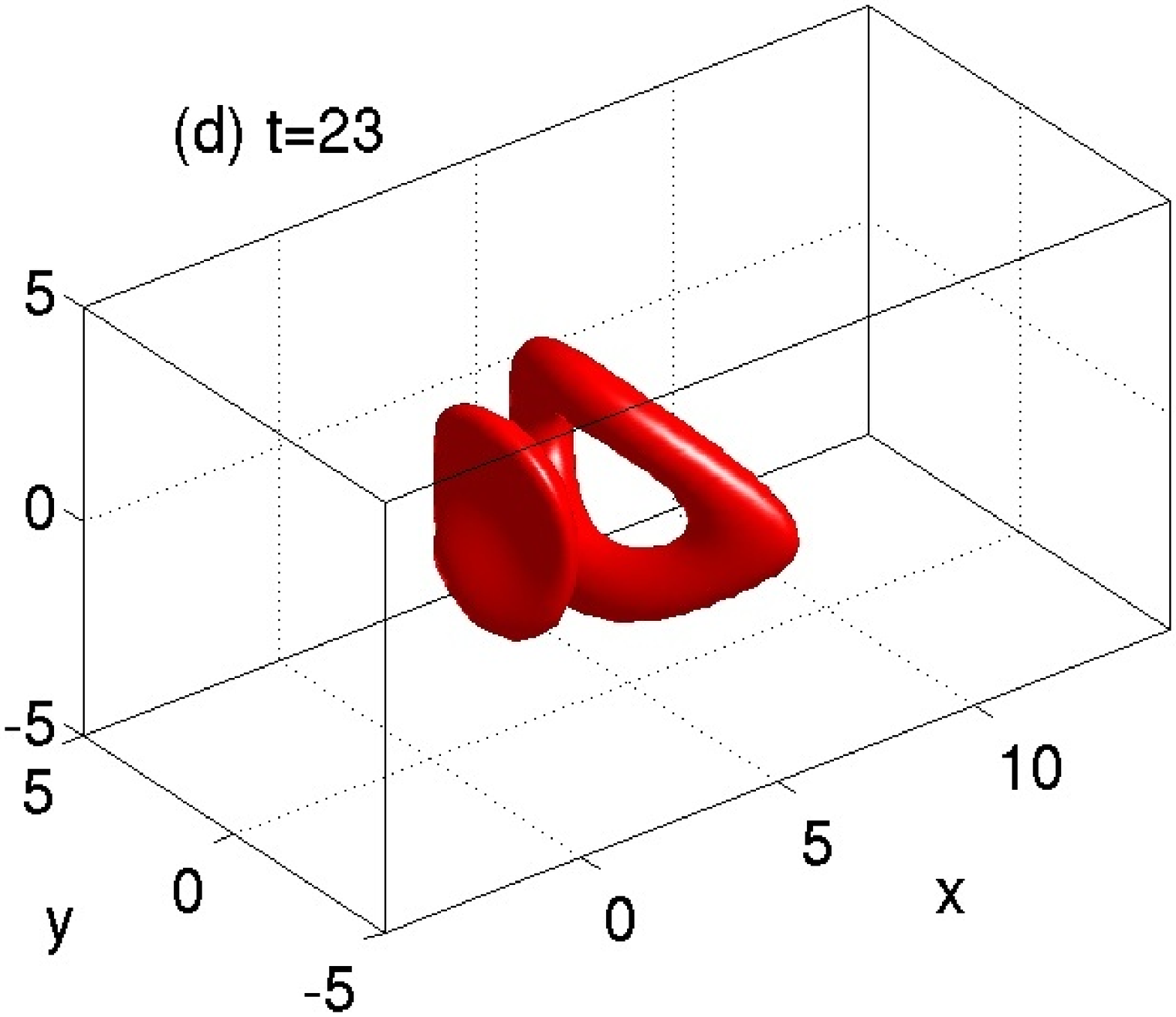}
\includegraphics[width=2.90cm]{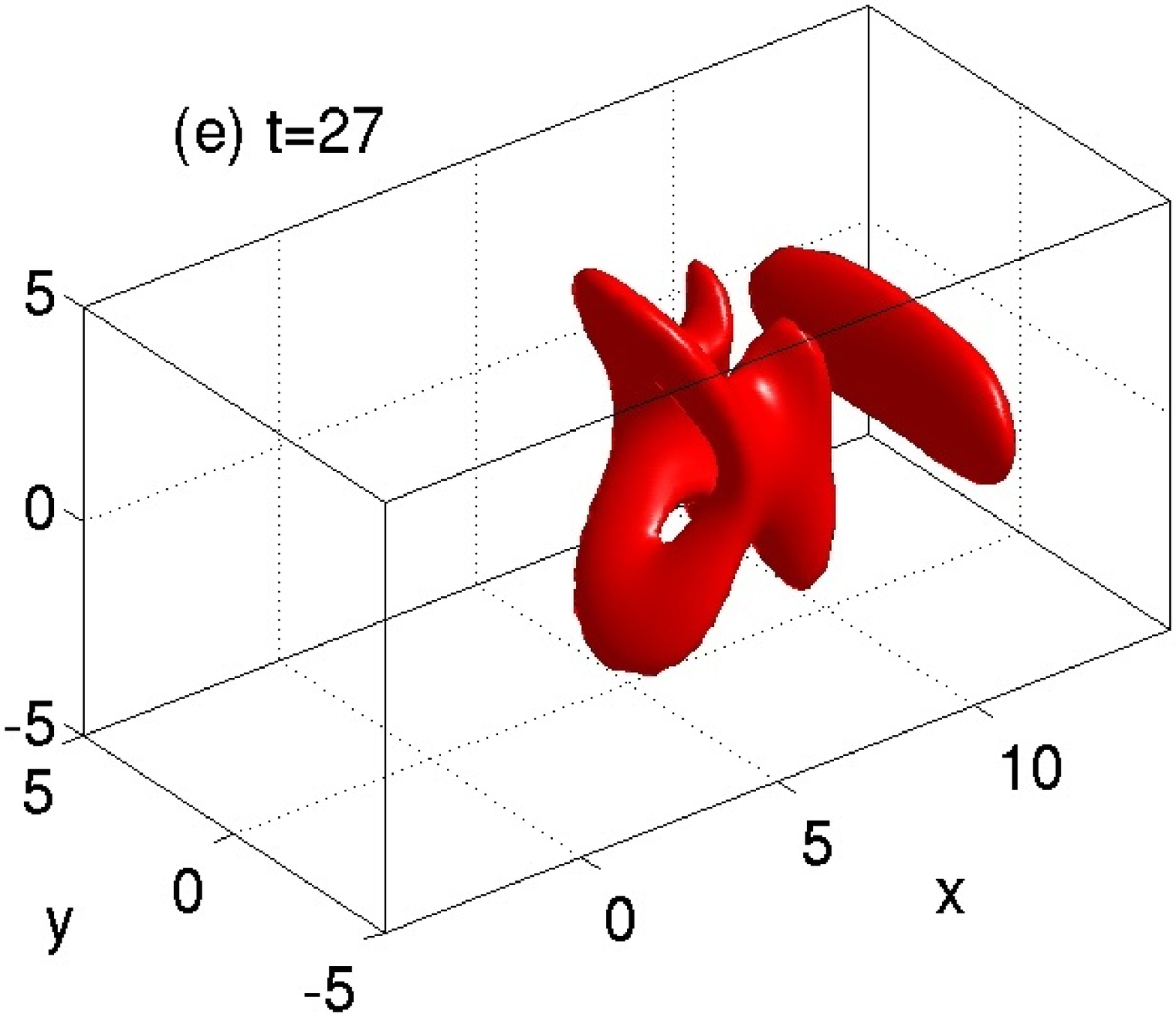}
\includegraphics[width=2.90cm]{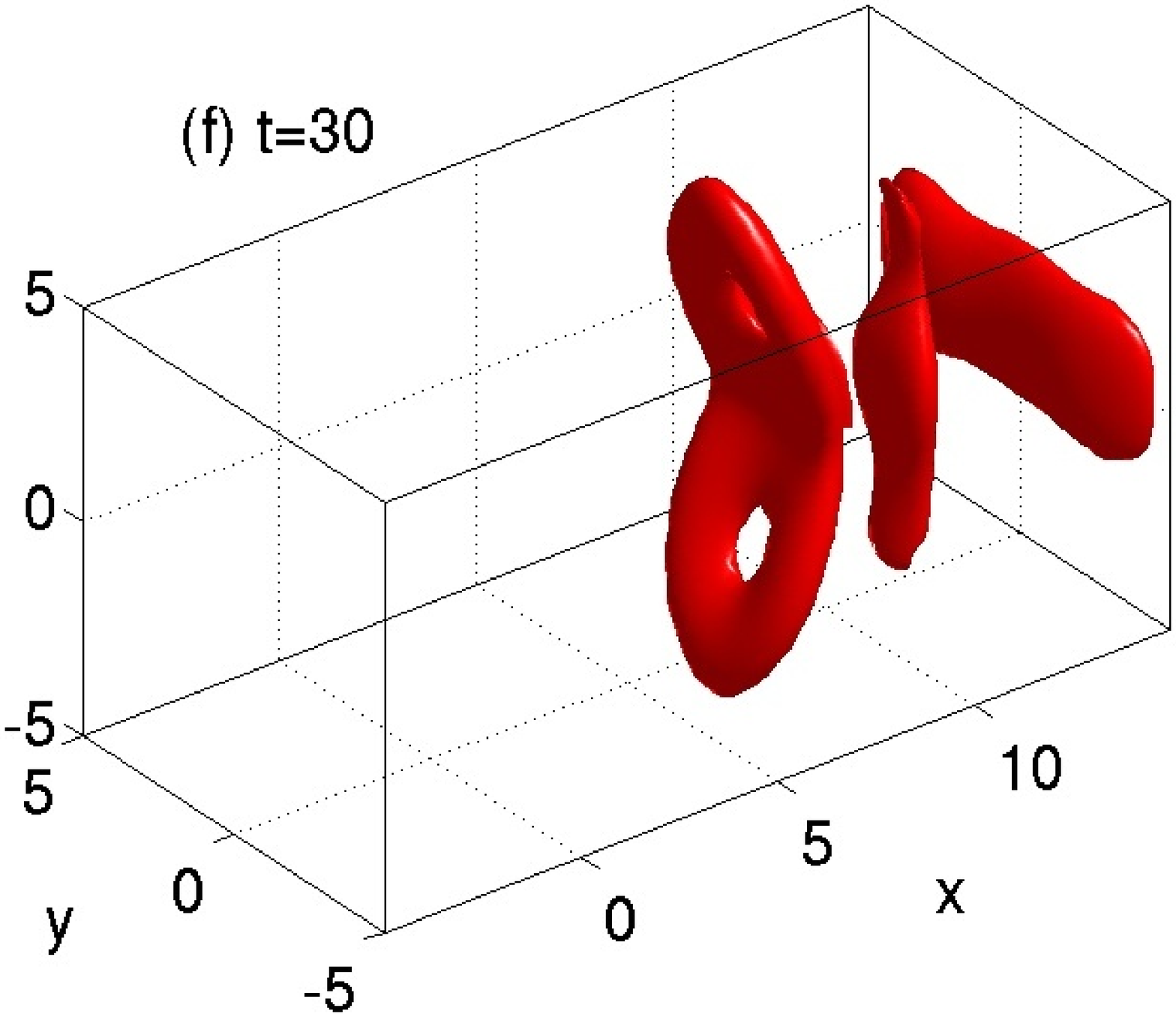}\\
\includegraphics[width=2.90cm]{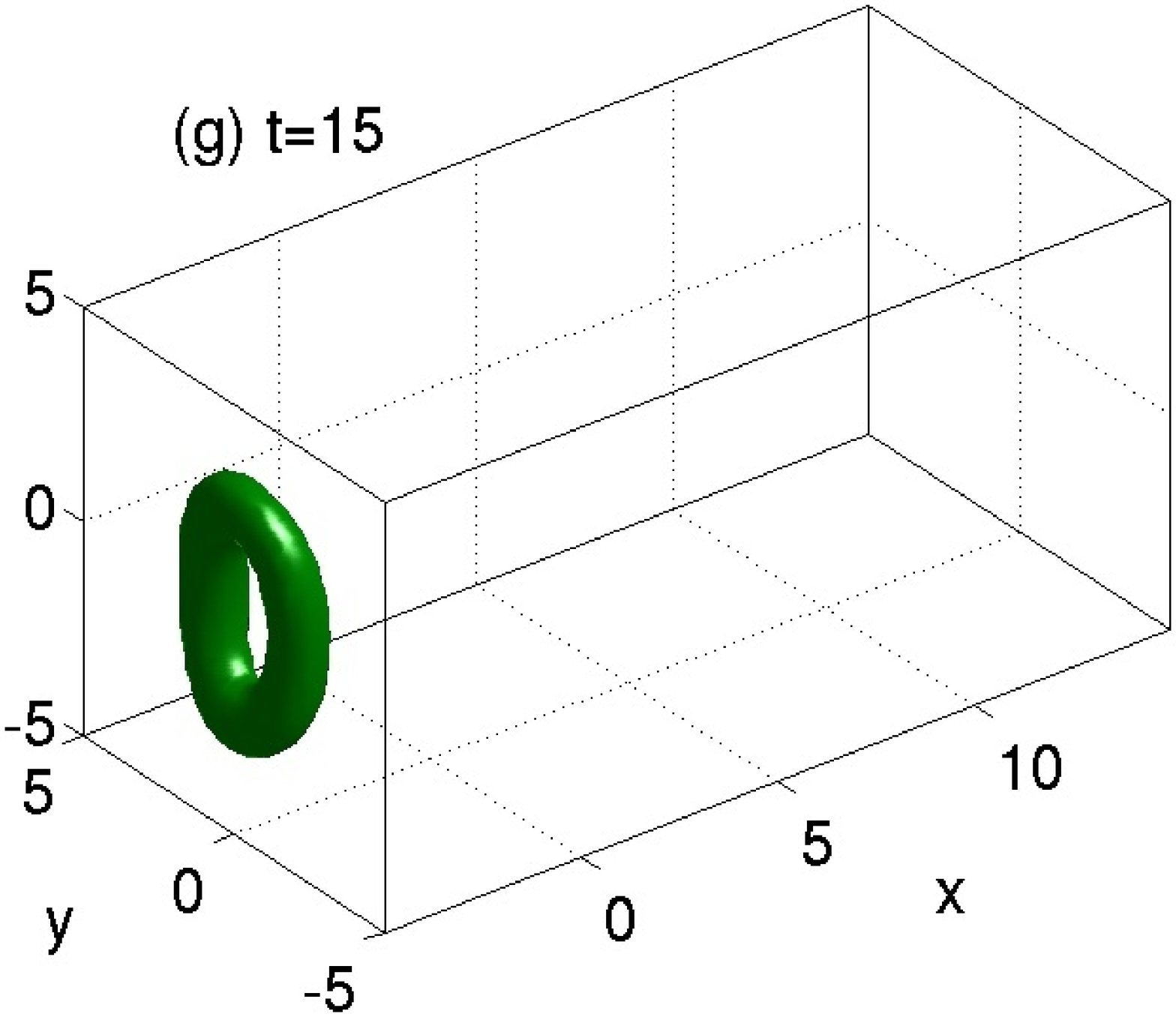}
\includegraphics[width=2.90cm]{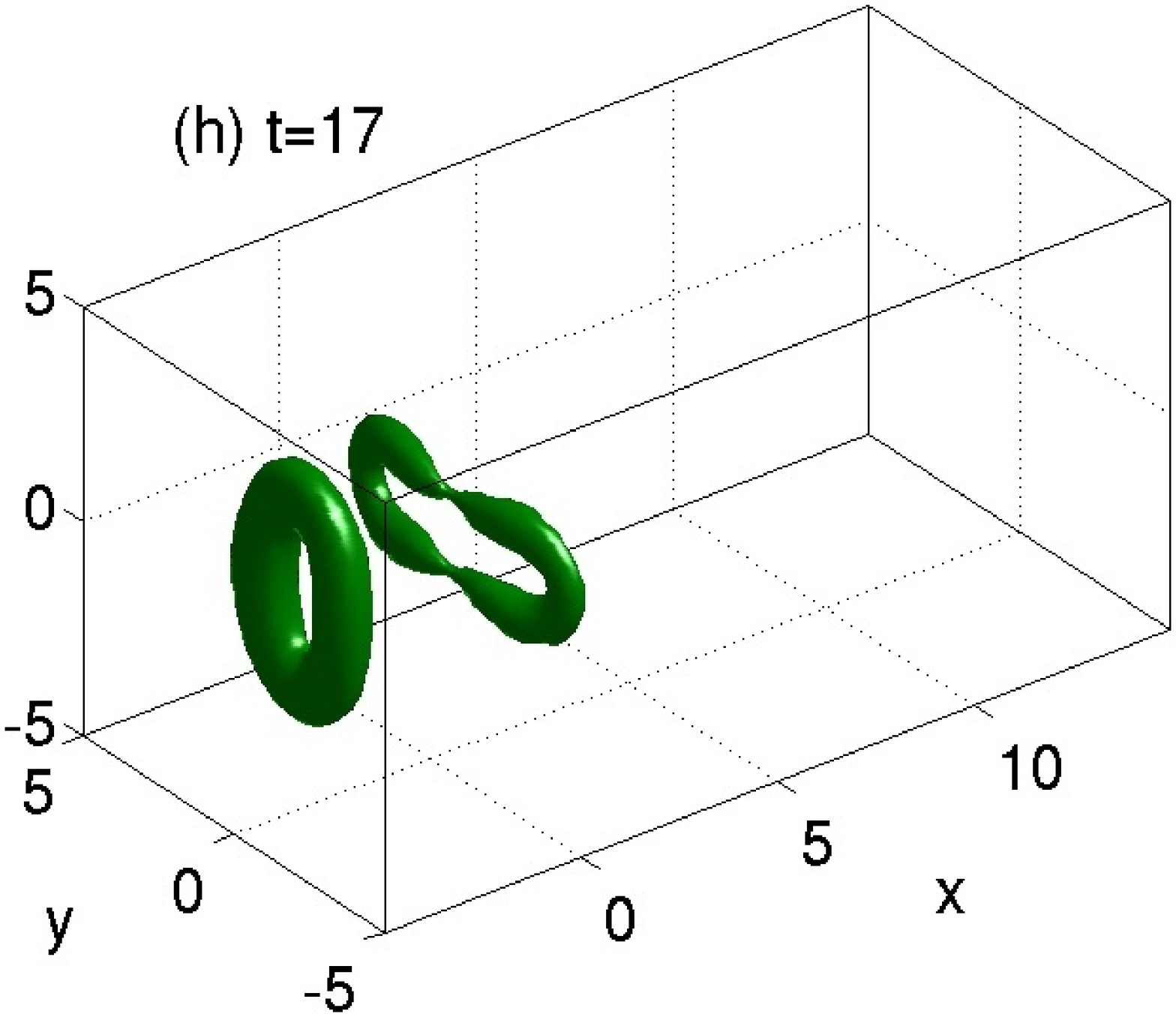}
\includegraphics[width=2.90cm]{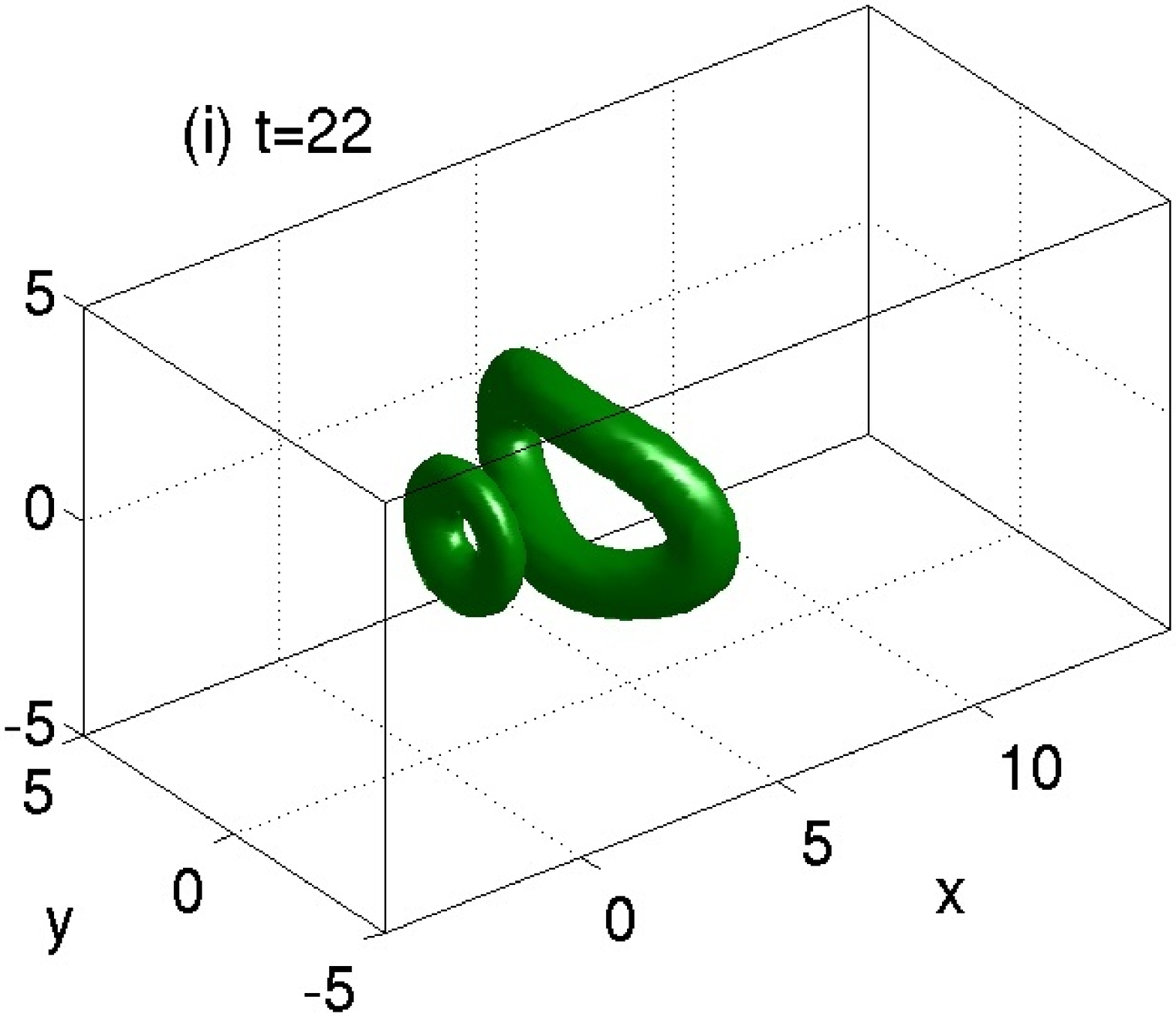}
\includegraphics[width=2.90cm]{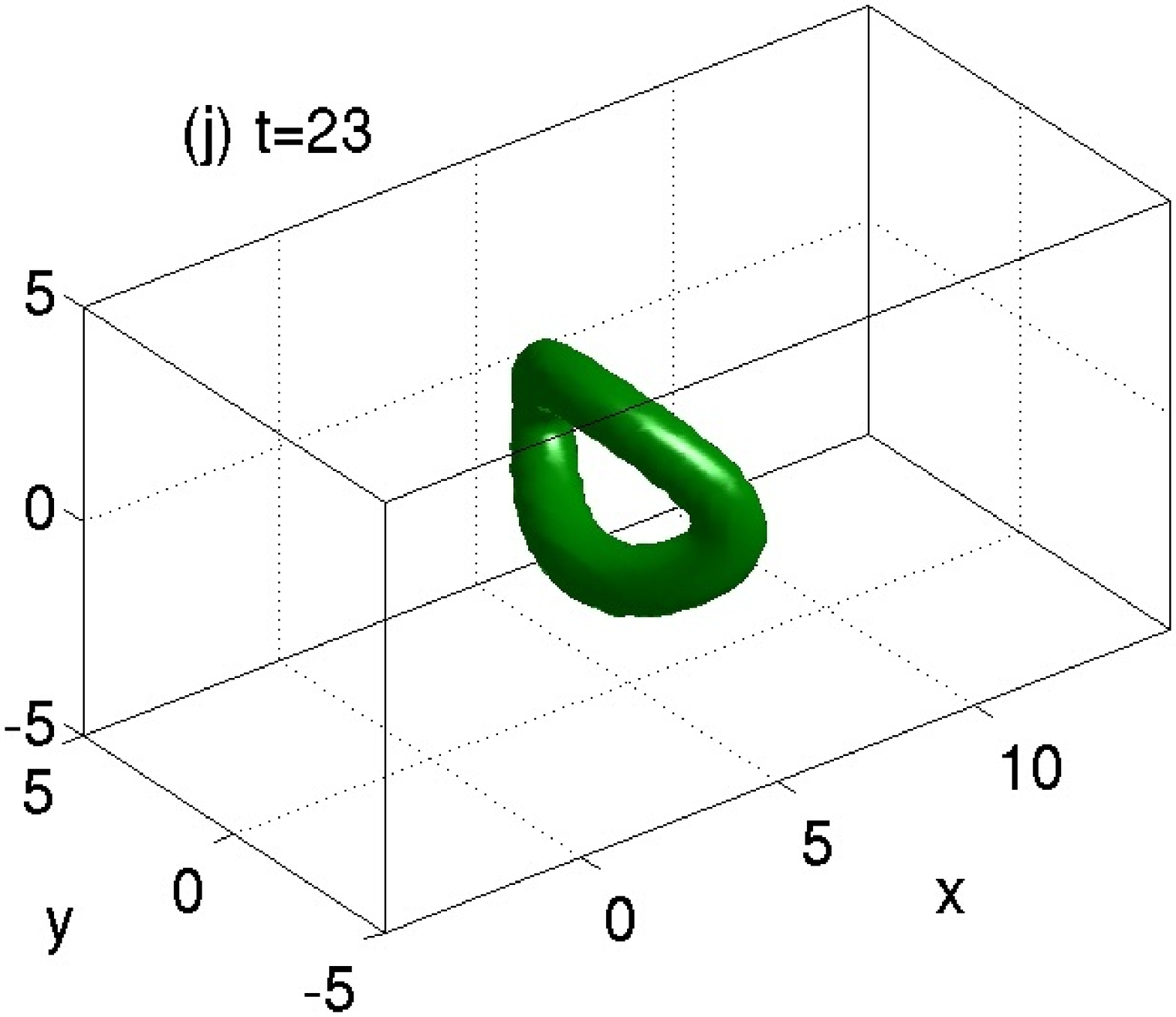}
\includegraphics[width=2.90cm]{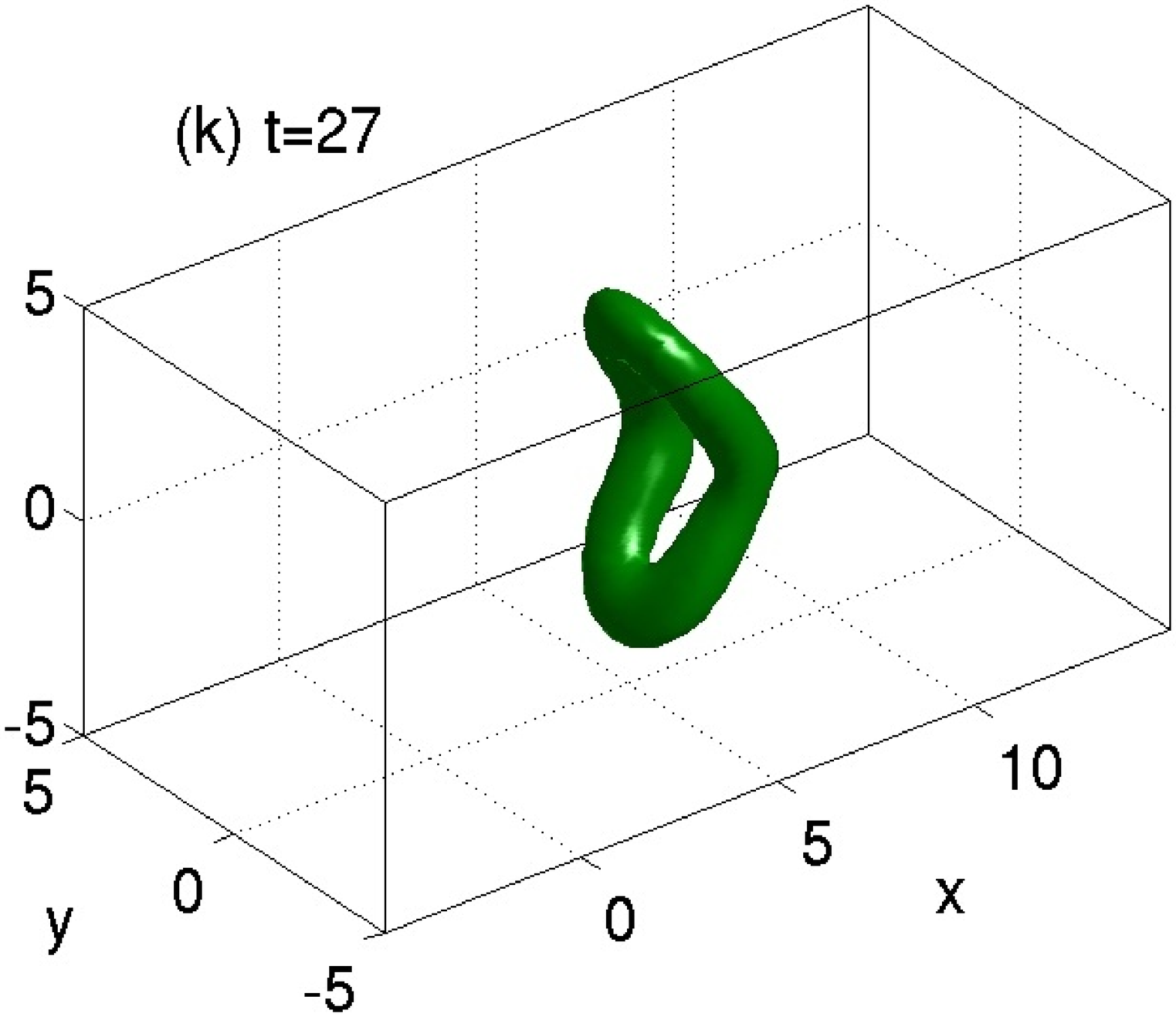}
\includegraphics[width=2.90cm]{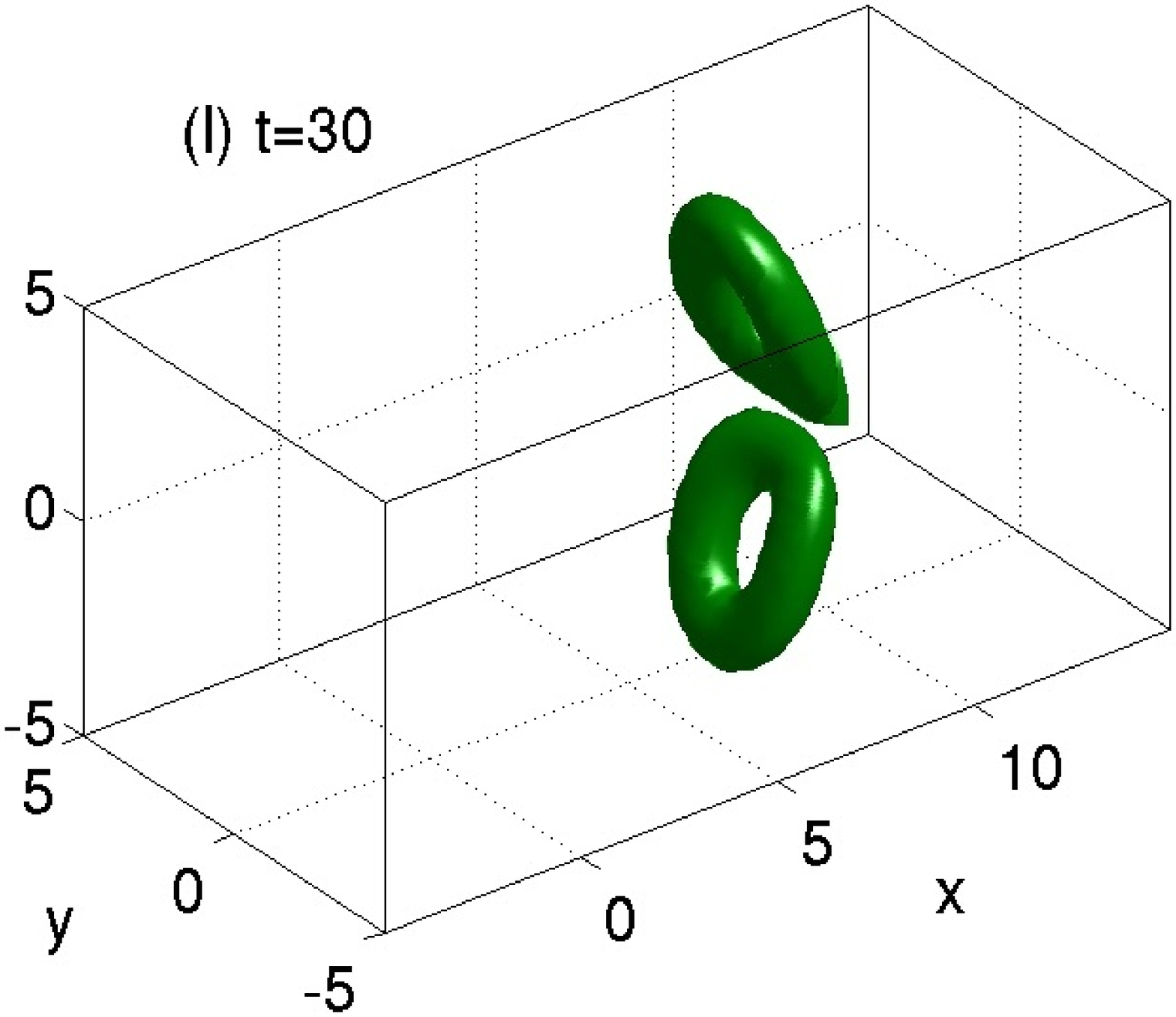}
\end{center}
\caption{(Color online)
Similar to Fig.~\ref{fig3Dw8vortvel08} but for a slightly larger
defect speed $s=1>c_2$. The top row of panels depicts the
iso-density contours of $\psi_0$ while the bottom row depicts the
respective vorticity isocontours. Note the successive nucleation of two
vortex rings. The first one shrinks and collapses into itself
between $t=22$ and $t=23$, while the second ring deforms and eventually
splits into two separate vortex rings between $t=29$ and $t=30$.
}
\label{fig3Dw8vortvel10}
\end{figure*}

\subsection{Three-dimensional setting}

Finally, we also performed 3D simulations, using a three-dimensional
generalization of the potential, elongated along the $z$-direction,
namely
\begin{eqnarray}
V&=&\frac{V_0}{16}\, \exp\left(-a (x-s t)^2\right)
%\label{sub3d}
\notag
\\&\times&
\left[\tanh\left(y+\frac{w_y}{2}\right)+1\right]
\left[\tanh\left(-y+\frac{w_y}{2}\right)+1\right]
\notag\\&\times&
\left[\tanh\left(z+\frac{w_z}{2}\right)+1\right]
\left[\tanh\left(-z+\frac{w_z}{2}\right)+1\right]
\notag
\end{eqnarray}
with $a=2$, $w_y=8$, $w_z=4$ and $V_0=0.9$. In Fig.~\ref{fig3Dw8vortvel08}
we depict the results for $s=0.8$ (i.e., above the second
critical speed). Panels (a)--(e) depict the iso-density contours
for $|\psi_0|^2$ at different times while panel (f) depicts
a superposition of the isocontours for the norm of the vorticity
field. As it can be seen from the figure, a vortex ring is formed in the
3D spinor condensate, as a result of the supercritical
nature of the chosen speed $s=0.8>c_2$.
The iso-density contours of $|\psi_0|^2$ clearly show
a depletion of atoms around the vortex ring that
is nucleated in the wake of the defect-induced region of
density minima.
It is worth stressing again that the dynamics of the different components
seems to be locked such that $|\psi_1|^2/A^2 \approx |\psi_0|^2/B^2
\approx |\psi_{-1}|^2/C^2$ and, therefore, we only depict
results for $\psi_0$ in the 3D case as well.

We also performed simulations for larger defect speeds
giving rise to a rich and complex scenario of multiple vortex rings
nucleations, collisions, collapses and splitting. A typical case
is shown in Fig.~\ref{fig3Dw8vortvel10} that corresponds
to the same parameters as in Fig.~\ref{fig3Dw8vortvel08}
but for a larger defect speed ($s=1$). The main characteristics
of the evolution can be summarized as follows: a first vortex
ring is nucleated in the wake of the defect at about $t=12$;
around $t=17$ a second vortex ring is nucleated while the
first vortex ring starts to shrink until it eventually disappears
around $t=23$; shortly after this, the second ring deforms and splits
into two separate vortex rings around $t=30$.

\section{Conclusions and Future Challenges}

We have studied the motion of a localized defect through a spinor $F=1$ condensate. 
Despite the three-component nature of the system, our systematic analysis of the 
small-amplitude excitation problem revealed that the nature of the nonlinearity is such that 
there appear not three, but merely two critical speeds in the system; these were identified 
analytically for the families of stationary uniform states of the system.
Our numerical simulations tested the dynamics for different values
of defect speeds in comparison to the two critical speeds.
Surprisingly, it was found that the lower one among
the two critical speeds is {\it not} activated and no
emission of nonlinear wave excitations emerges when this
threshold is crossed. On the other hand, when the defect
speed exceeds the second critical one, then emission of coherent structures 
arises independently of dimension; the resulting waveforms are dark solitons
in the one-dimensional setting, vortex-antivortex pairs in two dimensions, and
spinor vortex rings in the fully three-dimensional case.

While the present study showcases an experimentally accessible mechanism for producing nonlinear excitations in spinor BECs, a number of interesting questions are still outlying. In particular, perhaps the most relevant
question from a theoretical point of view involves acquiring a full understanding of why the first critical speed does not seem to be explored by the system, contrary to what might be
expected from the two-component case analyzed in Ref. \cite{susanto}.
Another direction of potential interest could be to explore in this
multi-component system what would happen if the speed of the defect becomes considerably
larger than the critical speed, in which case, and in the one-component
setting, a convective stabilization of oblique dark solitons
has been reported \cite{el}. Also interesting would be to study the formation of shock
waves \cite{chap01:dutton-soundsw-ablsw} in the spinor systems.
Work along these directions is currently in progress and will be reported in future publications.

\section*{Acknowledgments}

The authors (P.G.K, T.J.A., and Y.S.K.) acknowledge useful discussions with 
Elena Ostrovskaya. The work in Australia was supported by the Australian 
Research Council through the Australian Center for Quantum-Atom 
Optics (ACQAO) of the Australian Research Council. PGK and RCG
gratefully acknowledge support from NSF-DMS-0806762 and PGK also
acknowledges support from the NSF-CAREER program 
and the Alexander von Humboldt Foundation. The work of D.J.F. was 
partially supported by the Special Research Account of the University of Athens.
The work of A.S.R. was partially supported by FCT through grant
POCTI/FIS/56237/2004.

\end{document}